\documentclass[reprint,superscriptaddress,amsmath,amssymb,aps,pra,floatfix]{revtex4-1}
\usepackage[utf8]{inputenc}
\usepackage{amsmath,mathtools}
\usepackage{amssymb}
\usepackage{physics}
\usepackage{enumerate}
\usepackage{caption}
\usepackage{subcaption}
\usepackage{graphicx}
\usepackage{float}
\usepackage{mathtools}
\usepackage{bbold}
\usepackage{comment}
\usepackage{xcolor}

\DeclarePairedDelimiter\floor{\lfloor}{\rfloor}
\DeclarePairedDelimiter{\ceil}{\lceil}{\rceil}

\newcommand{\ba}{\begin{eqnarray}}
\newcommand{\ea}{\end{eqnarray}}
\newcommand{\ban}{\begin{eqnarray*}}
\newcommand{\ean}{\end{eqnarray*}}
\newcommand{\pguess}{P_\text{corr}}
\newcommand{\pguessFock}{\pguess^{(n)}}

\newcommand{\pud}{P_\text{UD}}
\newcommand{\nmax}{n_\text{max}}
\newcommand{\nbar}{\bar{n}}
\newcommand{\Sset}{\mathcal{S}}
\newcommand{\Oset}{\mathcal{O}}
\newcommand{\Rset}{\mathcal{R}}
\newcommand{\as}{\mathsf{a}}
\newcommand{\Nstate}{N}
\newcommand{\one}{\mathbb{1}}

\begin{document}

\title{Optimal single-shot discrimination of optical modes}

%\author{Chee Wei Soh}
%\affiliation{Department of Physics, National University of Singapore, 2 Science Drive 3, Singapore 117542, Singapore}

\author{Ignatius William Primaatmaja}
\affiliation{Centre for Quantum Technologies, National University of Singapore, 3 Science Drive 2, Singapore 117543, Singapore}

\author{Asaph Ho}
\affiliation{Centre for Quantum Technologies, National University of Singapore, 3 Science Drive 2, Singapore 117543, Singapore}

\author{Valerio Scarani}
\affiliation{Centre for Quantum Technologies, National University of Singapore, 3 Science Drive 2, Singapore 117543, Singapore}
\affiliation{Department of Physics, National University of Singapore, 2 Science Drive 3, Singapore 117542, Singapore}

\begin{abstract}
Retrieving classical information encoded in optical modes is at the heart of many quantum information processing tasks, especially in the field of quantum communication and sensing. Yet, despite its importance, the fundamental limits of optical mode discrimination have been studied only in few specific examples. Here we present a toolbox to find the optimal discrimination of any set of optical modes. The toolbox uses linear and semi-definite programming techniques, which provide rigorous (not heuristic) bounds, and which can be efficiently solved on standard computers. We study both probabilistic and unambiguous single-shot discrimination in two scenarios: the ``channel-discrimination scenario'', typical of metrology, in which the verifier holds the light source and can set up a reference frame for the phase; and the ``source-discrimination scenario'', more frequent in cryptography, in which the verifier only sees states that are diagonal in the photon-number basis. Our techniques are illustrated with several examples. Among the results, we find that, for many sets of modes, the optimal state for mode discrimination is a superposition or mixture of at most two number states; but this is not general, and we also exhibit counter-examples.

\end{abstract}

\maketitle

\section{Introduction}
In quantum optical systems, information can be encoded in the quantum state, in the optical mode, or in both \cite{treps}. In this paper, we consider situations in which \textit{classical information is encoded in optical modes} and address the problem of the ultimate limits in discriminating such modes (i.e.~in retrieving the classical information). Consider a set of $\Nstate$ modes associated to the annihilation operators $\{a_1, a_2,..., a_{\Nstate}\}$ and characterized by their commutation relations
\ba
[a_i, a_j^\dagger] = k_{ij} \one&\,,\,&\;|k_{ij}|\leq 1\,.
\label{eq:comm}
\ea
As any set of modes become distinguishable in the high intensity limit, we are interested in the \textit{single-shot discrimination of modes} under an energy constraint that fixes the average number of photons $\bar{n}$.

\begin{figure}[H]
    \centering
    \includegraphics[width= 1 \columnwidth]{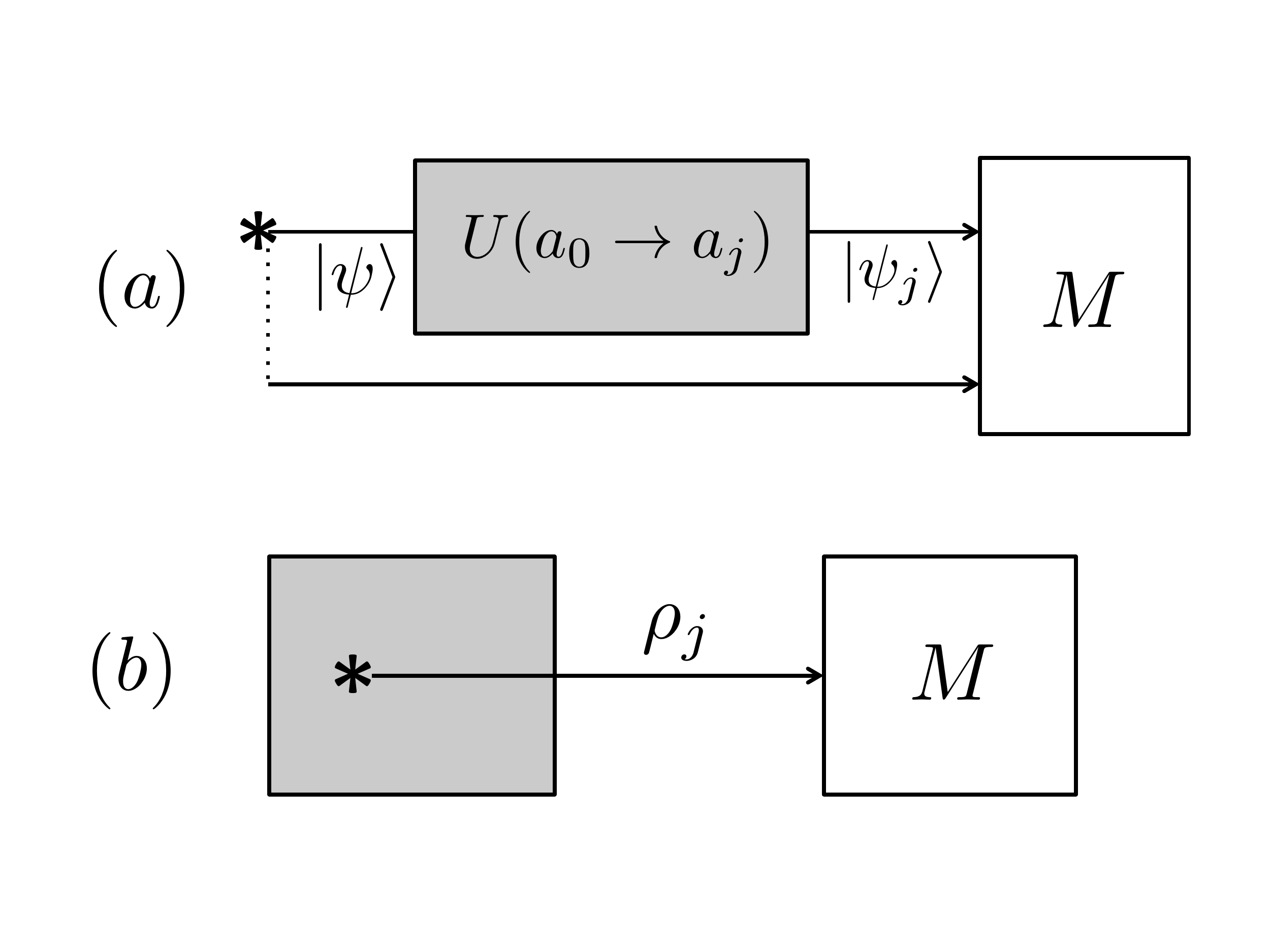}
    \caption{The two scenarios considered in this paper for mode discrimination. (a) Channel-discrimination scenario, studied in Section \ref{sec:channel}. The mode $a_j$ to be discriminated is encoded by a unitary transformation from an input mode $a_0$. The verifier has control of the source (star) and can add a reference beam, with respect to which the phase of the signal beam is well defined. If the reference is classical, the state in the signal beam can be taken as pure. (b) Source-discrimination scenario, studied in Section \ref{sec:source}. The source itself is in the black box, whence the signal exits encoded in the mode to be discriminated. For the verifier, the phase of the signal beam is global, and therefore the state appears as a photon-number mixture.}
    \label{fig0}
\end{figure}

Mode discrimination takes different forms, depending on the experimental scenario that is considered. We shall consider two scenarios in this paper. The \textit{channel-discrimination scenario} [Figure \ref{fig0} (a)], the mode is created by the unitary channel that maps a default mode $a_0$ onto one of the $a_j$. The source of light is in the hands of the verifiers, who can therefore avail themselves of a reference beam (``idler'') in addition to the beam that will be sent through the channel (``signal''). Then, the phase of the mode signal relative to the idler is defined: for instance it becomes possible to discriminate $a_1=a$ from $a_2=-a$. If the reference is classical (i.e.~an intense coherent state), as we shall assume here, the phase can be perfectly defined and the state in the channel can be taken to be pure; this choice is optimal for discrimination. The channel-discrimination scenario has a clearly \textit{metrological} flavor: besides the obvious case of phase discrimination \cite{banaszek2009,nair2012}, it can be seen as a special case of quantum reading \cite{pirandola11,nair,bisio,dallarno2012}, in which the devices to be discriminated are the unitary channels mentioned above. We refer to Ref.~\cite{pirandolareview} for a review of such metrological schemes.

The situation is different in the \textit{source-discrimination scenario}, in which the source of light is inside the same black box that performs the encoding of the mode, as sketched in panel (b) of Figure \ref{fig0}. From the point of view of the verifier, the phase of the signal mode is global and thus inaccessible. The state as seen from the verifier is the mixture over all possible values of the mode's phase, which is diagonal in the Fock basis \cite{molmer97} \footnote{This statement should not be misread as contradicting the known facts that successive pulses in a laser may have relative coherence \cite{vanenk}, and that such coherence may affect the unconditional security of cryptographic protocols, even if the encoding of classical information ignores those phases \cite{lopreskill}. For one, here the modes to be discriminated may include relative phases between physical ``pulses'' (see e.g.~Section \ref{sec: DPS}). Once these modes are decided, we are studying \textit{single-shot} discrimination, a task for which possible phases between successive instances of the modes indeed will not matter. In other words, we do not need to request the source to perform active phase randomization, for the state to be Fock-diagonal.}. The flavor of the source-discrimination scenario is more \textit{cryptographic}. Information is encoded in the modes in all the discrete-variable (e.g.~BB84 \cite{bennett1984}) and distributed-phase-reference (e.g.~DPS \cite{inoue2002differential}, COW \cite{stucki2005fast}) protocols for quantum key distribution (QKD); and also in protocols other than QKD, for instance quantum fingerprinting \cite{arrazola14,Jachura17}. For all these protocols, Eve ultimately wants to learn the mode in which the signals were encoded. Of course, the analysis of a cryptographic protocol goes beyond single-shot mode discrimination \cite{scarani09,xu2020secure,pirandola2019advances}. This may be the reason why, to the best of our knowledge, the latter has not been previously studied in the source-discrimination scenario.

In this paper, we provide recipes to compute \textit{upper bounds for single-shot discrimination (both probabilistic and unambiguous) of any set of modes} i.e. for arbitrary commutation relations \eqref{eq:comm}. Specifically, we show that those optimizations can be cast a semidefinite programming (SDP) relaxation based on the work of Ref.~\cite{wang2019characterising}. We present the recipes for the channel-discrimination scenario in Section \ref{sec:channel}, and that for the source-discrimination scenario in Section \ref{sec:source}. In Section \ref{sec:cases} we illustrate our method with several case studies: two modes (\ref{sec: two modes}), phase discrimination (\ref{sec: phase discrimination}), a family of $d$ modes and its Fourier-dual family (\ref{sec: FTbasis}), and the modes that appear in the DPS QKD protocol (\ref{sec: DPS}). Finally, in Section \ref{sec:losses} we discuss the extension of our study when losses are present between the device that encodes the modes and the measurement.

\section{Channel-discrimination scenario}
\label{sec:channel}
Let us first consider the channel-discrimination scenario. Suppose there are $\Nstate$ different optical modes $\mathcal{M} = \{a_1, ..., a_{\Nstate}\}$, with commutation relations given by Eq.~\eqref{eq:comm}. As mentioned previously, in the presence of reference beam, the phase of the signal mode could be defined relative to the the phase of the reference beam. Hence, the receiver's task is to discriminate between pure states $\Rset = \{ \ket{\psi_1}, \ket{\psi_2}, ..., \ket{\psi_{\Nstate}} \}$ where
\begin{equation} \label{eq: pure states}
    \ket{\psi_j} = \sum_{n = 0}^\infty c_n \frac{(a_j^\dagger)^n}{\sqrt{n!}} \ket{0} \equiv \sum_{n = 0}^\infty c_n \ket{n_j}
\end{equation}
where $c_n$ is an arbitrary complex number such that $\abs{c_n}^2 = p_n$ is the probability of the source emitting $n$ photons. We also introduced the shorthand $\ket{n_j}$ denoting the $n$-photon state in the mode $a_j$. The inner-product of the states associated to different modes can be computed easily
\begin{equation} \label{eq: inner product}
    \braket{\psi_i}{\psi_j} = \sum_{n=0}^\infty p_n k_{ij}^n
\end{equation}
Note that the inner-product depends only on the photon number distribution $\{p_n\}$ and the commutation relations between different mode defined in Eq.~\eqref{eq:comm}.

\subsection{Probabilistic mode discrimination}

We first consider the setting for probabilistic discrimination. For uniform priors (our method can be easily generalised to any fixed priors), the optimal guessing probability is given by
\begin{equation}
    \pguess^{\text{opt}} = \max_{\{M_j\},\{c_n\}}\frac{1}{\Nstate} \sum_{j = 1}^{\Nstate} \bra{\psi_j} M_j \ket{\psi_j}
\end{equation}
where $M_j$ is the POVM element associated with the outcome $j$ and $c_n$ is the coefficient of the state when written in the photon number basis. In other words, we have to optimize both the state and the measurements that maximize the guessing probability, subject to the energy constraint. Let us now cast this optimization into a SDP by adapting the techniques of \cite{wang2019characterising}. Before getting to it, we notice that \cite{wang2019characterising,navascues2007bounding,navascues2008convergent} considers a hierarchy of semidefinite relaxations, which in general only yield upper bounds on the guessing probability. However, since we only consider a single receiver with no classical inputs, going into the second level of the hierarchy will satisfy the rank loop condition \cite{navascues2008convergent} and hence the first level of the hierarchy is actually tight.

Here comes the construction. Consider the set $\Oset = \{\one, M_1, ..., M_{\Nstate} \}$. As discussed in \cite{wang2019characterising, navascues2007bounding, navascues2008convergent}, the $\{M_i\}$ can be taken as projective measurements without any loss of generality. Denoting $O_i$ the elements of $\Oset$, one can define a set of vectors $\Sset= \{O_i \ket{\psi_j}: O_i \in \Oset, \ket{\psi_j} \in \Rset \}$. Since all Gram matrices are positive semidefinite (PSD), so is the $\Nstate(\Nstate+1)\times \Nstate(\Nstate+1)$ Gram matrix $G$ associated to the set $\Sset$. Hence we have
\begin{equation} \label{eq: SDP_prob_ideal}
\begin{split}
\pguess^{\textrm{opt}}=  &\max_{G, \{p_n\}}\frac{1}{\Nstate} \sum_{j = 1}^{\Nstate} \bra{\psi_j} M_j \ket{\psi_j}\\
\text{s.t.} \  & p_n \geq 0 \qquad \forall n \\
& \sum_{n} p_n = 1\\
& \sum_{n} p_n n = \nbar\\
& G \succeq 0\\
& \braket{\psi_i}{\psi_j}=\sum_{n}p_n k_{ij}^n \qquad \forall i,j\,,
\end{split}
\end{equation}
where the last relation is Eq.~\eqref{eq: inner product} and determines the entries of $G$ associated with $O_0=\one$.

However, this is an optimization problem with infinitely many variables $ p_n$ and hence is computationally intractable. We then relax it by truncating the number of photons to $\nmax$ (i.e.~we'll have $\nmax+1$ variables $p_n$). We do it in such a way as to obtain an \textit{upper bound} $\pguess^{\textrm{SDP}}\geq \pguess^{\text{opt}}$ on the mode discrimination probabilities: that is, we derive some necessary (but not sufficient) conditions on the photon number distribution, and as a result the feasible region may be larger than the one allowed by quantum theory. Clearly, the relaxation can be made arbitrarily tight by increasing the photon number cut-off; and we expect $\pguess^{\text{opt}}\approx \pguess^{\textrm{SDP}}$ when $\bar{n}\ll \nmax$.

We then define the truncated state
\begin{equation}\label{truncform}
    \ket{\tilde{\psi}_j} = \sum_{n = 0}^{\nmax} c_n \frac{1}{\sqrt{n!}} {a_j^\dagger}^n \ket{0}
\end{equation}
that is sub-normalised:
\begin{equation}
    \sum_{n = 0}^{\nmax} p_n \leq 1\,.
\end{equation}
The inner-product of the truncated states is
\begin{equation}
    \braket{\tilde{\psi_i}}{\tilde{\psi_j}} = \sum_{n = 0}^{\nmax} p_n k_{ij}^n\label{truncinner}
\end{equation}
and its difference with the inner product of the full states can be bounded as
\begin{equation}
\begin{split}
        \abs{\braket{\tilde{\psi_i}}{\tilde{\psi_j}} - \braket{\psi_i}{\psi_j}} &=  \abs{\sum_{n > \nmax} p_n k_{ij}^n} \\ &\leq \sum_{n > \nmax} p_n \abs{k_{ij}}^n 
        \\ &\leq \left(1 -  \sum_{n = 0}^{\nmax} p_n \right) \abs{k_{ij}}^{\nmax + 1} \equiv \varepsilon_{ij}\label{trunceps}
\end{split}
\end{equation}
where the first inequality is a consequence of triangle inequality and the second inequality is due to the fact that $\abs{k_{ij}} \leq 1$. The constraint on the mean photon number can be relaxed to
\begin{equation}
    \sum_{n = 0}^{\nmax} p_n n + \left(\nmax + 1 \right) \left(1 - \sum_{n = 0}^{\nmax} p_n \right) \leq \nbar\,.
\end{equation}
All in all, we have $\pguess^{\text{opt}}\leq \pguess^{\textrm{SDP}}$ with
\begin{equation} \label{eq: SDP_probabilistic}
\begin{split}
\pguess^{\textrm{SDP}}=  &\max_{G, \{p_n\}}\frac{1}{\Nstate} \sum_{j = 1}^{\Nstate} \bra{\psi_j} M_j \ket{\psi_j}\\
\text{s.t.} \  &G \succeq 0\\
& p_n \geq 0 \qquad \forall n \leq \nmax\\
& \sum_{n \leq \nmax} p_n \leq 1\\
& \sum_{n \leq \nmax} p_n (\nmax + 1 - n) \geq \nmax + 1 - \nbar\\
& \abs{\braket{\psi_i}{\psi_j}-\braket{\tilde{\psi_i}}{\tilde{\psi_j}}} \leq \varepsilon_{ij} \qquad \forall i,j
\end{split}
\end{equation}
where the last constraint uses the expressions \eqref{truncinner} and \eqref{trunceps}. That constraint is linear in the $p_n$ when $k_{ij}$ is real; when $k_{ij}$ is complex, we can re-write it as
\begin{equation}
    \begin{pmatrix}
    \varepsilon_{ij} & \braket{\psi_i}{\psi_j}-\braket{\tilde{\psi_i}}{\tilde{\psi_j}} \\
    \left(\braket{\psi_i}{\psi_j} -\braket{\tilde{\psi_i}}{\tilde{\psi_j}}\right)^* &\varepsilon_{ij}
    \end{pmatrix} \succeq 0
\end{equation}
which is a matrix inequality linear in the $p_n$, hence a valid SDP constraint.

\subsection{Unambiguous mode discrimination}

The same technique can be adapted to find the maximum success probability for unambiguous mode discrimination. For unambiguous discrimination, one must allow for the inconclusive outcome, which we associate to the POVM element $M_\varnothing$. The Gram matrix $G$ will correspondingly increase in size to $\Nstate(\Nstate+2)\times\Nstate(\Nstate+2)$. Then we must add the constraint that the probability of error is zero
\begin{equation}
    \sum_{j = 1}^{\Nstate} \sum_{i \neq j, i \neq \varnothing} \bra{\psi_j} M_i \ket{\psi_j} = 0
\end{equation}
In the study of unambiguous discrimination, the aim is to minimize the probability of the inconclusive outcome or, equivalently, to maximize the success probability. Assuming uniform priors, the success probability is given by
\begin{equation}
    \pud = 1 - \frac{1}{\Nstate} \sum_{j=}^{\Nstate} \bra{\psi_j} M_\varnothing \ket{\psi_j}
\end{equation}
Note that both the success probability and the error probability are linear functions of the Gram matrix $G$. Hence, putting everything together, we have $\pud^\text{opt} \leq \pud^{\textrm{SDP}}$ with
\begin{equation} \label{eq: SDP_UMD}
\begin{split}
\pud^{\textrm{SDP}}= &\max_{G, \{p_n\}} 1 - \frac{1}{\Nstate} \sum_{j=1}^{\Nstate} \bra{\psi_j} M_\varnothing \ket{\psi_j}\\
\text{s.t.} \  & p_n \geq 0 \qquad \forall n \leq \nmax\\
& \sum_{n \leq \nmax} p_n \leq 1\\
& \sum_{n \leq \nmax} p_n (\nmax + 1 - n) \geq \nmax + 1 - \nbar\\
& G \succeq 0\\
& \abs{\braket{\psi_i}{\psi_j}-\braket{\tilde{\psi_i}}{\tilde{\psi_j}}} \leq \varepsilon_{ij} \qquad \forall i,j \\
& \sum_{j = 1}^{\Nstate} \sum_{i \neq j, i \neq \varnothing} \bra{\psi_j} M_i \ket{\psi_j} = 0
\end{split}
\end{equation}

Recall that unambiguous state discrimination is possible if and only if the states to be discriminated are linearly independent. But unambiguous mode discrimination is possible even for linearly dependent modes, as the linear independence of the states is provided by the multiphoton components. In fact, the families studied in subsections \ref{sec: phase discrimination}, \ref{sec: FTbasis} and \ref{sec: DPS} will be linearly dependent.

\section{Source-discrimination scenario}
\label{sec:source}
In the source-discrimination scenario, the task is to discriminate between states $\{\rho_1, \rho_2, ..., \rho_{\Nstate}\}$ that are \textit{diagonal in the Fock basis}, subject to the energy constraint. Indeed, even assuming that the source produces a pure state
\begin{equation} \label{eq: pure states2}
    \ket{\psi_j(\theta)} = \sum_{n = 0}^\infty c_n \frac{(e^{i\theta}a_j^\dagger)^n}{\sqrt{n!}} \ket{0}
\end{equation}
(this is Eq.~\eqref{eq: pure states} with explicit mention of the global phase $\theta$ of the signal mode), in the absence of a reference beam, the information available to the receiver is the phase-randomized state
\ba
\rho_j&=&\int_0^{2\pi}\frac{d\theta}{2\pi}\ket{\psi_j(\theta)}\bra{\psi_j(\theta)}\,=\,\sum_{n = 0}^\infty p_n \ketbra{n_j}{n_j}\,.\label{eq: mixed state}
\ea
Since the states to be discriminated are diagonal in the Fock basis, nothing is lost if the receiver starts by measuring the number of photons, then uses the best discrimination strategy for the given value of $n$. This will manifest itself in the possibility of splitting the optimization in two steps.

\subsection{Probabilistic mode discrimination}

For probabilistic mode discrimination, the guessing probability is given by
\begin{align}
    \pguess^{\text{opt}} &= \max_{\{p_n\}, \{M_j\}} \frac{1}{\Nstate} \sum_{j=1}^{\Nstate} \Tr(\rho_j M_j) %\nonumber\\
    %&= \max_{\{p_n\}} \sum_{n = 0}^\infty p_n \left[ \frac{1}{N}\max_{\{M_j\}} \sum_{j=1}^{\Nstate}  \bra{n_j} M_j \ket{n_j} \right]
    \nonumber\\&
    \equiv \max_{\{p_n\}} \sum_{n = 0}^\infty p_n \pguessFock \label{eq: pguess mixed state}
\end{align}
where
\begin{equation}
    \pguessFock = \max_{\{M_j\}} \frac{1}{\Nstate} \sum_{j=1}^{\Nstate} \bra{n_j} M_j \ket{n_j} \label{eq: pguess Fock state}
\end{equation} is the optimal guessing probability for $n$ photons. Therefore, as expected, we can split the optimization into two steps. In the first step, we solve \eqref{eq: pguess Fock state} for each value of $n$. This can be done using the SDP technique of Ref.~\cite{wang2019characterising}. Similar to what we have done in the channel-discrimination scenario, consider the set $\Rset_n = \{\ket{n_1}, \ket{n_2}, ... , \ket{n_\Nstate} \}$ which are Fock state $n$ from the set of modes $\mathcal{M}$. We define the set of vectors $\Sset_n = \{O_i\ket{n_j}: O_i \in \Oset, \ket{n_j} \in \Rset_n\}$. Now, denote the Gram matrix associated to $\Sset_n$ by $G^{(n)}$. The SDP to bound $\pguessFock$ is
\begin{equation} \label{eq: pguess Fock}
    \begin{split}
        \pguessFock = &\max_{G^{(n)}} \frac{1}{\Nstate} \sum_{j=1}^{\Nstate} \bra{n_j} M_j \ket{n_j}\\
        & \text{s.t. } \ G^{(n)} \succeq 0\\
        & \qquad \braket{n_i}{n_j} = k_{ij}^n
    \end{split}
\end{equation}
which we have to solve for each photon number $n$. In the second step, having $\pguessFock$ for each photon number $n$, we just need to enforce the energy constraint. This remaining step is linear programming (LP):
\begin{equation} \label{eq: LP pg exact}
\begin{split}
    \pguess^{\text{opt}} = &\max_{\{p_n\}} \sum_n p_n \pguessFock\\
    &\text{s.t. } \ p_n \geq 0 \qquad \forall n,\\
    & \qquad \sum_{n} p_n = 1, \\
    & \qquad \sum_{n} p_n n = \nbar
\end{split}
\end{equation}
Like in the channel-discrimination scenario, we have infinitely many variables $p_n$, so we need a cutoff that relaxes the original LP. With the same arguments as above, the resulting relaxation $\pguess^\text{LP}\geq \pguess^\text{opt}$ is given by
\begin{equation} \label{eq: LP pguess relaxed}
\begin{split}
    \pguess^{\text{LP}} = &\max_{\{p_n\}} \sum_{n = 0}^{\nmax} p_n \pguessFock + \left(1 - \sum_{n=0}^{\nmax} p_n \right) \\
    &\text{s.t. } \ p_n \geq 0 \qquad \forall n \leq \nmax,\\
    & \qquad \sum_{n = 0}^{\nmax} p_n \leq 1, \\
    & \qquad \sum_{n=0}^{\nmax} p_n n + \left(1 - \sum_{n=0}^{\nmax} p_n\right) (\nmax+1) \leq \nbar
\end{split}
\end{equation}
Notice that this relaxation is equivalent to assuming that the modes are perfectly distinguishable for $n > \nmax$. This is a good approximation when we pick sufficiently high photon number cutoff $\nmax$.

\subsection{Unambiguous mode discrimination}

%\begin{equation}
%    \sum_{j = 1}^{\Nstate} \sum_{i \neq j, i \neq \varnothing} \bra{n_j} M_i \ket{n_j} = 0
%\end{equation}
%for any photon number $n$.

Also for unambiguous state discrimination the optimal success probability is of the form
\begin{equation}
    \pud^\text{opt} = \max_{\{p_n\}}  \sum_{n} p_n \pud^{(n)}
\end{equation}
and can be bounded in two steps. In the first step, $\pud^{(n)}$ is computed from the SDP
\begin{equation} \label{eq: pud Fock}
    \begin{split}
        \pud^{(n)} = &\max_{G^{(n)}} 1 - \frac{1}{\Nstate} \sum_{j=1}^{\Nstate} \bra{n_j} M_\varnothing \ket{n_j}\\
        & \text{s.t. } \ G^{(n)} \succeq 0\\
        & \qquad \braket{n_i}{n_j} = k_{ij}^n\\
        & \qquad \bra{n_j} M_i \ket{n_j} = 0, \quad \forall i \neq j, i \neq \varnothing
    \end{split}
\end{equation}
where the last constraint captures the unambiguous discrimination condition \begin{equation}
    \sum_{j = 1}^{\Nstate} \sum_{i \neq j, i \neq \varnothing} \Tr(\rho_j M_i) = 0\,,
\end{equation}
which is indeed satisfied if and only if $\bra{n_j} M_i \ket{n_j} = 0$ for all $n$ whenever $i \neq \varnothing$, $j \neq i$. In the second step, the energy constraint is enforced in a LP, which we write down directly for the relaxation $\pud^{\text{LP}}\geq \pud^{\text{opt}}$ with a photon-number cutoff:
\begin{equation} \label{eq: LP pud relaxed}
\begin{split}
    \pud^{\text{LP}} = &\max_{\{p_n\}} \sum_{n = 0}^{\nmax} p_n \pud^{(n)} + \left(1 - \sum_{n=0}^{\nmax} p_n \right) \\
    &\text{s.t. } \ p_n \geq 0 \qquad \forall n \leq \nmax,\\
    & \qquad \sum_{n = 0}^{\nmax} p_n \leq 1, \\
    & \qquad \sum_{n=0}^{\nmax} p_n n + \left(1 - \sum_{n=0}^{\nmax} p_n\right) (\nmax+1) \leq \nbar\,.
\end{split}
\end{equation}

\subsection{Towards an analytical solution of the LP}
\label{ss: solveLP}

We have just seen that the final step of the optimization for the source-discrimination scenario is a LP of the form
\begin{equation} \label{eq: LP pguess exact}
\begin{split}
    P^{\text{opt}} = &\max_{\{p_n\}} \sum_n p_n \as_n\\
    &\text{s.t. } \ p_n \geq 0 \qquad \forall n,\\
    & \qquad \sum_{n} p_n = 1, \\
    & \qquad \sum_{n} p_n n = \nbar
\end{split}
\end{equation} where $\as_n=\pguessFock$ for probabilistic discrimination and $\as_n=\pud^{(n)}$ for unambiguous discrimination. For a given $\as_n$, this LP can be solved analytically. We are going to show how this can be done, and highlight a condition on $\as_n$ under which the solution can be easily spelled out.

The LP \eqref{eq: LP pguess exact} is written in the so-called \textit{primal} form. One could also consider its \textit{dual} form given by \cite{boyd2004convex}
\begin{equation}\label{eq: LP pguess dual}
    \begin{split}
    d^* = \min_{x,y} \ & x + \nbar y \\
    \text{s.t.} \,\, & y \geq -\frac{1}{n} \left(x - \as_n \right)\quad \forall n\\
    \end{split}
\end{equation}
Due to the strong duality of LP, we know that $d^* = P^\text{opt}$ and hence it is sufficient to solve the dual problem.

\begin{figure}[H]
    \centering
    \includegraphics[width=\columnwidth]{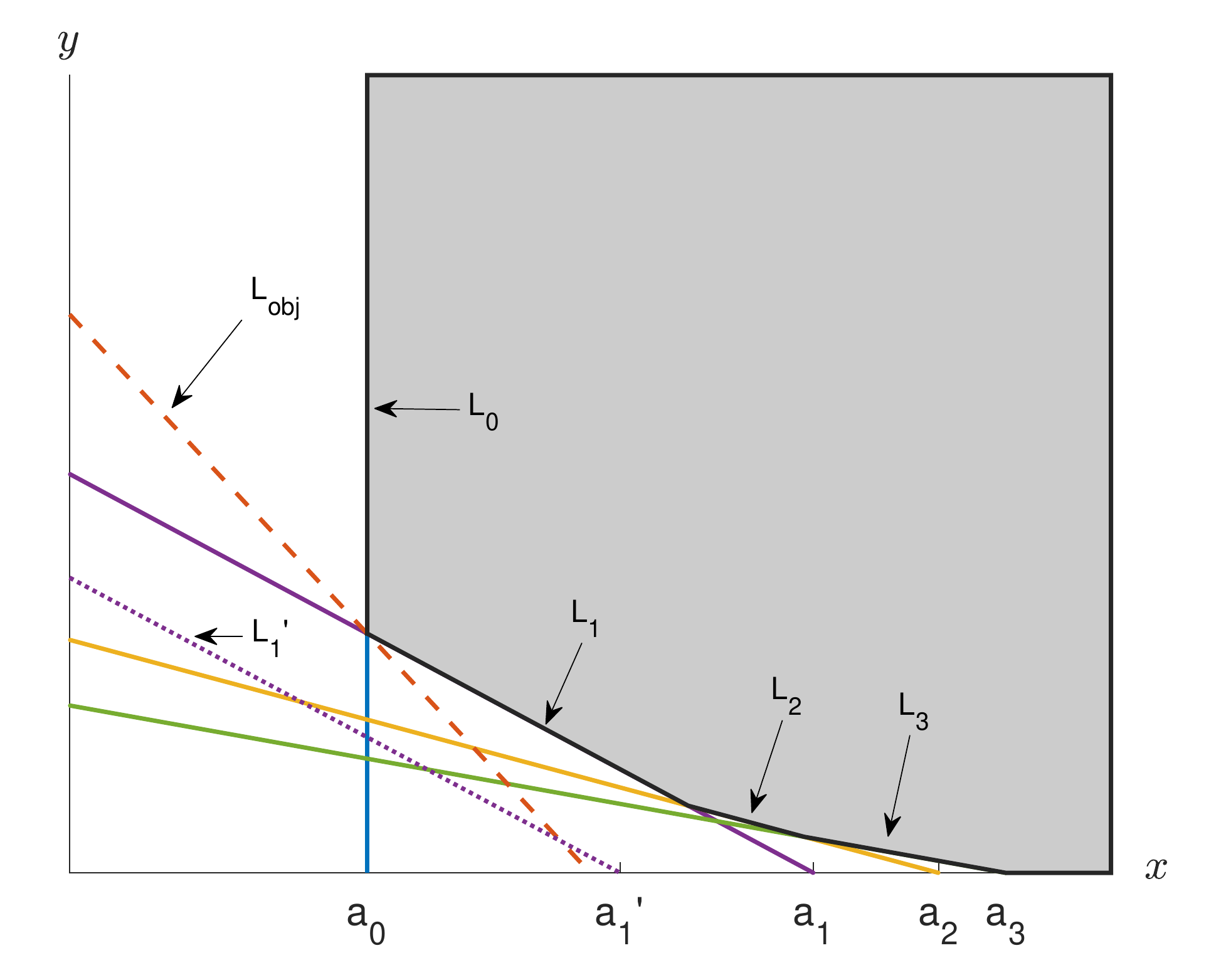}
    \caption{Geometrical solution of the dual LP \eqref{eq: LP pguess dual}. The first constraint is associated to the line $L_0: x = \as_0$, and is satisfied in the half-space $x \geq \as_0$. For $n > 0$, the constraint is associated to the line $L_n: y = -\frac{1}{n} (x - \as_n)$ and the region satisfying that constraint is the upper half-space above $L_n$. The \textit{feasible region} is the region where all constraints are satisfied (shaded area, drawn for $n\leq 3$). The objective function is a line $L_\text{obj}$ with gradient $-1/\nbar$ (dashed; for the plot, $\nbar = 0.5$). The dual problem can hence be understood as finding the minimum $y$-intercept by translating the line $L_\text{obj}$ vertically while ensuring that it still touches the feasible region. The $\as_n$ satisfy \eqref{eq: condition} and thus all the $L_n$ contribute non-trivially to the boundary of the feasible region. Had we chosen $\as_1'$ instead of $\as_1$, the line $L_1'$ (dotted) would not contribute to that boundary; that is, the constraint for $n=1$ would be always satisfied within the feasible region.}
    \label{fig: feasible region}
\end{figure}

Whereas the primal problem is an optimization over infinitely many variables with a few equality constraints, the dual problem is an optimization over two variables with infinitely many constraints, which define the \textit{feasible region}. The way to solve the dual problem is easily understood geometrically 
(Fig.~\ref{fig: feasible region}). One traslates the lines $L_{\text{obj}}(d)=\{(x,y)\;|\,x+\nbar y=d\}$ with fixed gradient $-1/\nbar$, till finding the lowest one that has at least one point in the feasible region. But the boundary of the feasible region is given by segments of straight lines $L_n$ of gradient $-1/n$. So the limiting line $L_{\text{obj}}(d^*)$ may touch the boundary of the feasible set either in a single point (the intersection of two $L_n$, as illustrated in the figure) or in a whole segment. The latter can only happen if $\nbar=n$, but this is not the only condition: it is further necessary that $L_n$ contributes to the boundary of the feasible region. This may not always be the case, as illustrated in Fig.~\ref{fig: feasible region}. By this recipe, one can always find the solution, given the $\as_n$.

It is worth describing in detail the case where \textit{all} the constraints in \eqref{eq: LP pguess dual}, i.e.~all the $L_n$, contribute to the boundary of the feasible region in a non-trivial way. Given that the gradient of $L_n$ is $-1/n$, a necessary and sufficient condition for the boundary to be as described is that $x_{n-1,n}<x_{n,n+1}$ where $(x_{n,m}, y_{n,m})$ are the coordinates of the intersection of $L_n$ with $L_m$. From $y_{n-1,n} =-\frac{1}{n-1} (x_{n-1,n} -\as_{n-1})= -\frac{1}{n} (x_{n-1,n} -\as_{n})$ one immediately finds $x_{n-1,n} = n \as_{n-1} - (n-1) \as_n$. Thus, $x_{n-1,n}<x_{n,n+1}$ will be the case if and only if 
\begin{equation} \label{eq: condition}
    \as_{n-1} - 2 \as_n + \as_{n+1} < 0\,.
\end{equation}

If this condition is satisfied, then the optimal discrimination takes up a very clear form. Indeed, for a boundary as described:
\begin{itemize}
\item[(i)] If $\nbar\not\in\mathbb{N}$, the intersection that defines $d^*$ will be with a single point, namely the intersection of $L_{\floor{\nbar}}$ and $L_{\ceil{\nbar}}$. The optimal state is then a mixture of two Fock states with these numbers, and the suitable weights.
\item[(ii)] If $\nbar=n\in\mathbb{N}$, the intersection that defines $d^*$ will be the whole segment of gradient $-1/n$, and the optimal state with be the Fock state $\rho=\ket{n}\bra{n}$.
\end{itemize}
In other words, the solution of the LP will be $P^\textrm{opt}=p_{\floor{\nbar}}\as_{\floor{\nbar}}+p_{\ceil{\nbar}}\as_{\ceil{\nbar}}$, with
\begin{equation}\label{floorceil}
    p_n =
    \begin{cases}
    1 + \floor{\nbar} - \nbar & \text{if $n = \floor{\nbar}$,}\\
    \nbar - \floor{\nbar} & \text{if $n = \floor{\nbar} + 1$,}\\
    0 &\text{otherwise}\,.
    \end{cases}
\end{equation} Thus, in many cases we can expect the optimal state for discrimination to consist of the Fock state $\ket{\nbar}$ if $\nbar\in\mathbb{N}$, and of the suitable mixture of the Fock states $\ket{\floor{\nbar}}$ and $\ket{\ceil{\nbar}}$ if $\nbar\not\in\mathbb{N}$. However, condition \eqref{eq: condition} does not always hold: in subsection \ref{sec: FTbasis} we shall see an example where it is not met for $n=1$, and indeed the Fock state $\ket{1}$ will not be optimal for $\nbar = 1$.

%As elaborated in Fig.~\ref{fig: feasible region}, finding the optimal solution for the dual problem is equivalent to translating the line $L_{\text{obj}}$ vertically along the feasible region. Using simple geometric argument, the optimal solution must be obtained in either of the following two scenarios: when $L_\text{obj}$ passes through an intersection of two different $L_n$ (when $\nbar$ is not an integer) or when $L_\text{obj}$ coincides with one of the $L_n$ (when $\nbar$ is an integer and hence $L_\text{obj}$ will be parallel to one of the $L_n$).

\section{Case studies}
\label{sec:cases}

We have derived efficient relaxations for estimating the parameters of mode discrimination, both probabilistic and unambiguous, in both the channel-discrimination and the source-discrimination scenario. In this Section, we discuss some case studies. 

\subsection{Two modes}
\label{sec: two modes}

%The discrimination between two optical modes has been the object of previous studies \cite{pirandola11,nair, bisio, dallarno2012}, and the results that we derive here could be inferred from those works. We present it here in an original, self-contained way. \textbf{VS: this is probably the case for one scenario, not for the other. I'll check it.}

The discrimination between two modes is determined by a single parameter
\begin{equation}
    [a_1, a_2^\dagger] = k \one\;,\quad k\in\mathbb{C}\,,\;|k|\leq 1\,.
\end{equation} When $k = 1$, the two modes are identical and therefore indistinguishable. When $k = 0$, the two modes are orthogonal and can be perfectly distinguished when $\nbar\geq 1$.

Besides using our numerical tools, we are going to derive analytical solutions, exploiting the fact that the discrimination of two equally probable pure states has been solved long ago for both probabilistic \cite{helstrom1969quantum} and unambiguous discrimination  \cite{ivanovic87,dieks88,peres88}:
\begin{equation}\label{helstrom}
    \begin{split}
\pguess^{\text{opt}}&=\frac{1}{2}(1+\sqrt{1-\abs{\langle \psi_1|\psi_2\rangle}^2})\\
\pud^{\text{opt}}&=1-\abs{\langle \psi_1|\psi_2\rangle}\,.
\end{split}
\end{equation}

With this, single-shot discrimination of two modes in the \textit{source-discrimination scenario} can be fully solved analytically. Indeed, there is no need to solve the SDPs \eqref{eq: pguess Fock} and \eqref{eq: pud Fock} since we know from \eqref{helstrom} that $\pguess^{(n)}=\frac{1}{2}(1+\sqrt{1-\abs{k}^{2n}})$ and $\pud^{(n)}=1-\abs{k}^{n}$ (notice that everything depends on $\abs{k}$.). Moving to the LP, it is immediate to verify that both expressions satisfy condition \eqref{eq: condition}. So we can import from subsection \ref{ss: solveLP} that the solution is \begin{equation}\label{eq:sd2}
P_{\textrm{corr/UD}}^\textrm{opt}=p_{\floor{\nbar}}P_{\textrm{corr/UD}}^{(\floor{\nbar})}+p_{\ceil{\nbar}}P_{\textrm{corr/UD}}^{(\ceil{\nbar})}\end{equation} with the $p_n$ given in \eqref{floorceil}.

In the \textit{channel-discrimination scenario}, we know from \eqref{helstrom} that $\pguess^{\text{opt}}=\frac{1}{2}(1+\sqrt{1-\chi^2})$ and $\pud^{\text{opt}}=1-\chi$ with \begin{equation} \label{eq: minscal}
    \begin{split}
        \chi = &\min_{\{p_n\}}  \left|\sum_{n=0}^{\infty}p_nk^n \right|\\
        &\text{s.t. } \ p_n \geq 0 \qquad \forall n,\\
    & \qquad \sum_{n} p_n = 1, \\
    & \qquad \sum_{n} p_n n = \nbar
    \end{split}
\end{equation} where we used the expression \eqref{eq: inner product} of the scalar product. Instead of the SDPs, we could try and solve \eqref{eq: minscal}.

For $k\geq 0$, it is a LP of the form \eqref{eq: LP pguess dual} with $\as_n\equiv -k^n$ (notice that here we are minimizing, whence the sign). Condition \eqref{eq: condition} reads $-k^{n-1}(1+k^2)<0$ and is therefore satisfied: so we know that the solution is
\ba
\chi=p_{\floor{\nbar}}k^{\floor{\nbar}}+p_{\ceil{\nbar}}k^{\ceil{\nbar}}&\quad&[k\geq 0]\label{eq: chi}
\ea with the $p_n$ given in \eqref{floorceil}. The corresponding optimal state is $\sqrt{p_{\floor{\nbar}}}\ket{\floor{\nbar}}+e^{i\varphi}\sqrt{p_{\ceil{\nbar}}}\ket{\ceil{\nbar}}$ for any $\varphi$. This value of $\chi$ shows that $\pguess^{\textrm{opt}}$ is larger than in the source-discrimination scenario \eqref{eq:sd2}, since $\sqrt{1-\chi^2}\geq p_{\floor{\nbar}}\sqrt{1-k^{2\floor{\nbar}}}+p_{\ceil{\nbar}}\sqrt{1-k^{2\ceil{\nbar}}}$; while the value of $\pud^{\textrm{opt}}$ is identical in the two scenarios. 

For $k<0$, the optimization \eqref{eq: minscal} is also LP; but whether the absolute value adds a minus sign or not (i.e.~whether $\as_n=+k^n$ or $-k^n$) is not known \textit{a priori}. One would therefore have to solve the two LPs, then compare the solutions. In either case, condition \eqref{eq: condition} would be satisfied only for alternate $n$, and so the solution is not expected to involve only $\floor{\nbar}$ and $\ceil{\nbar}$. As for $k\in\mathbb{C}\setminus\mathbb{R}$, the optimization \eqref{eq: minscal} is quadratic, and there is no guarantee that an analytical solution can be found.

\begin{figure}[H] 
    \centering
    \begin{subfigure}[b]{\columnwidth}
        \includegraphics[width= \columnwidth]{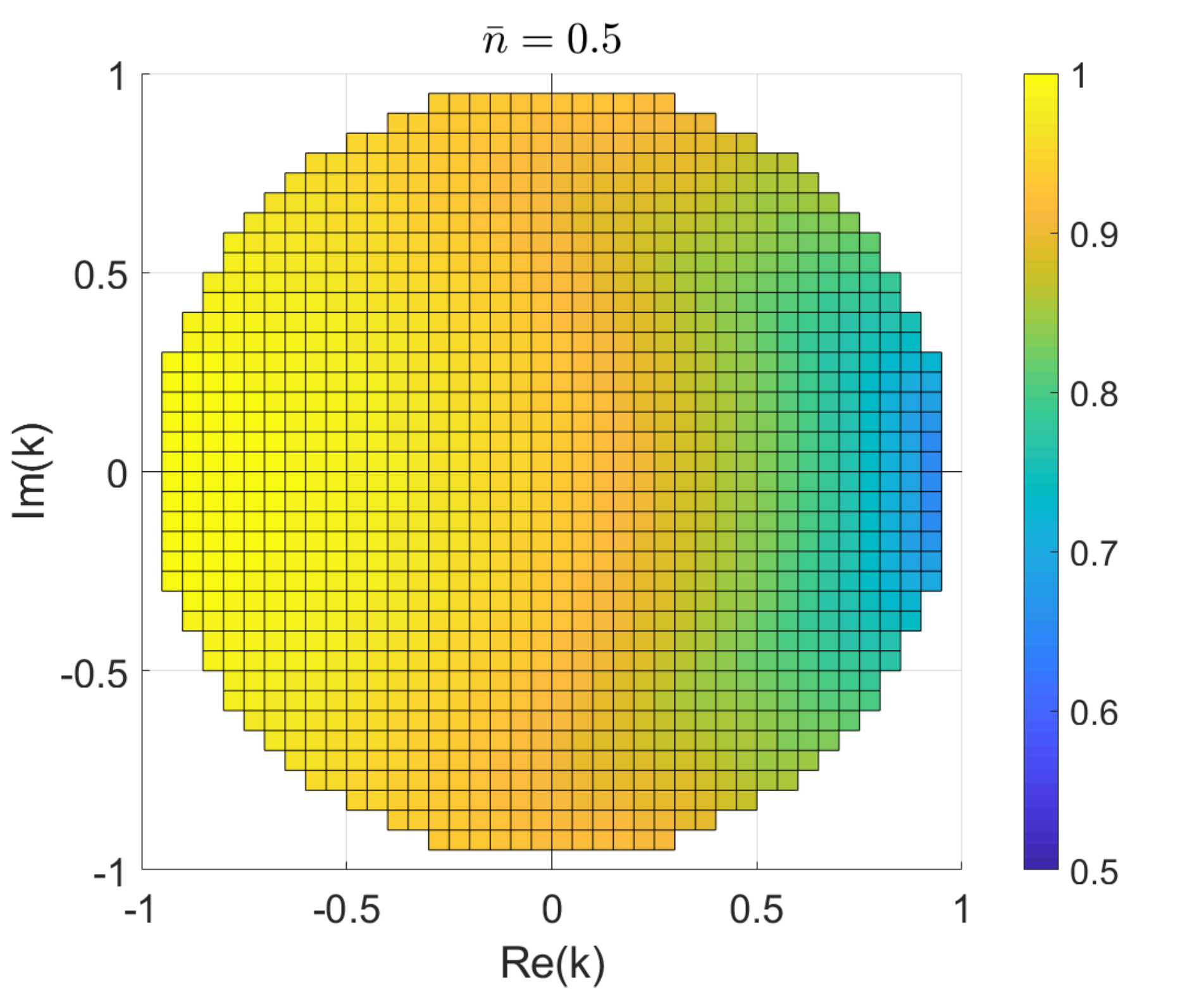}
        \caption{Upper bound $\pguess^\text{SDP}$ for $\nbar = 0.5$}
        \label{fig: two_mode_nbar=0.5}
    \end{subfigure}
    \begin{subfigure}[b]{\columnwidth}
        \includegraphics[width= \columnwidth]{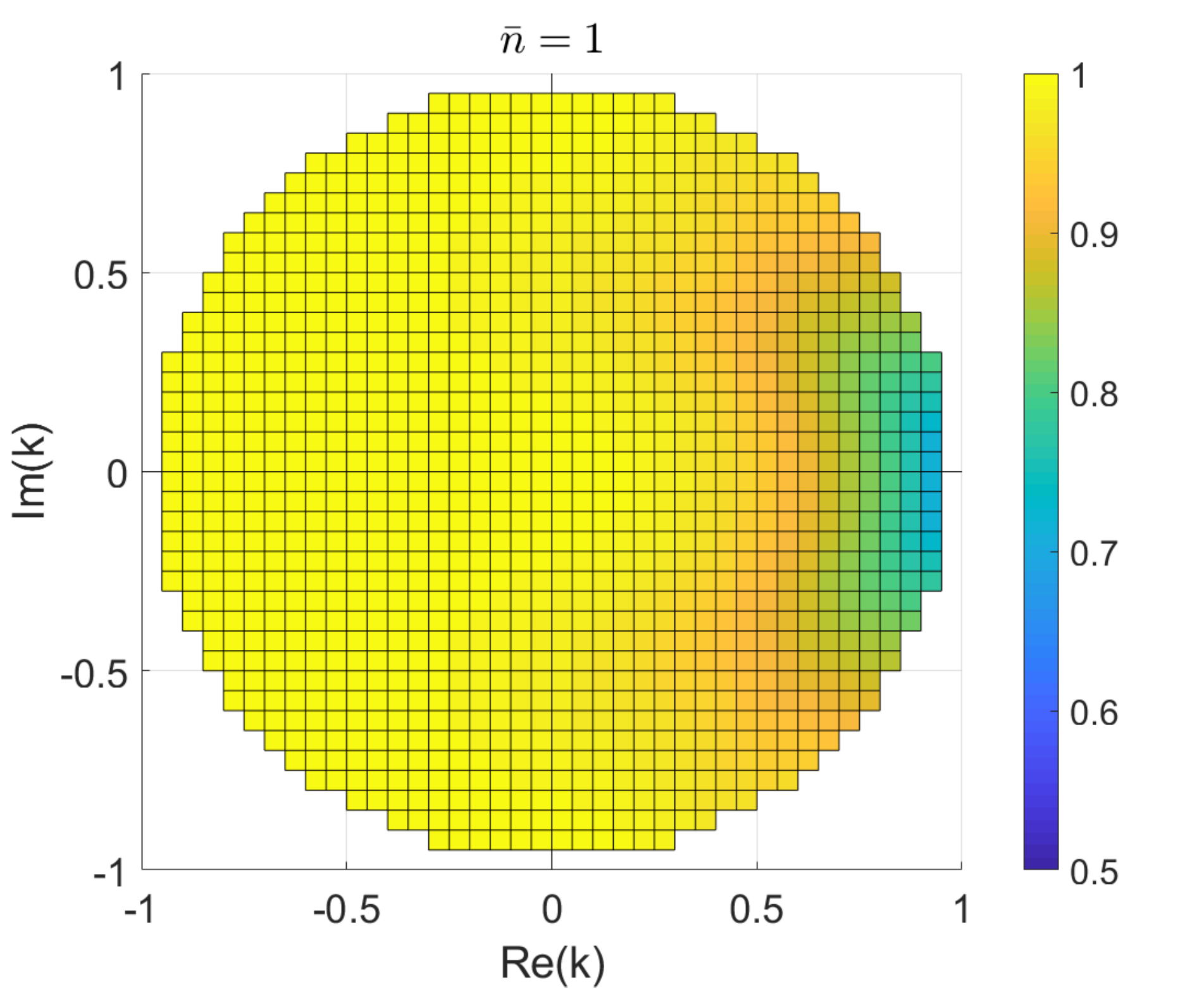}
        \caption{Upper bound $\pguess^\text{SDP}$ for $\nbar = 1$}
        \label{fig: two_mode_nbar=1}
    \end{subfigure}
    \caption{\textbf{Probabilistic channel-discrimination between two modes: dependence on the mode overlap $k\in\mathbb{C}$}. Upper bound $\pguess^\text{SDP}$ on the guessing probability, solution of the SDP \eqref{eq: SDP_probabilistic} for $\nmax = 300$, in a polar plot of $k$, for $\nbar = 0.5$ and $\nbar = 1$.} \label{fig: two_mode_prob}
\end{figure}

\begin{figure}[H]
    \centering
    \begin{subfigure}[b]{\columnwidth}
        \includegraphics[width= \columnwidth]{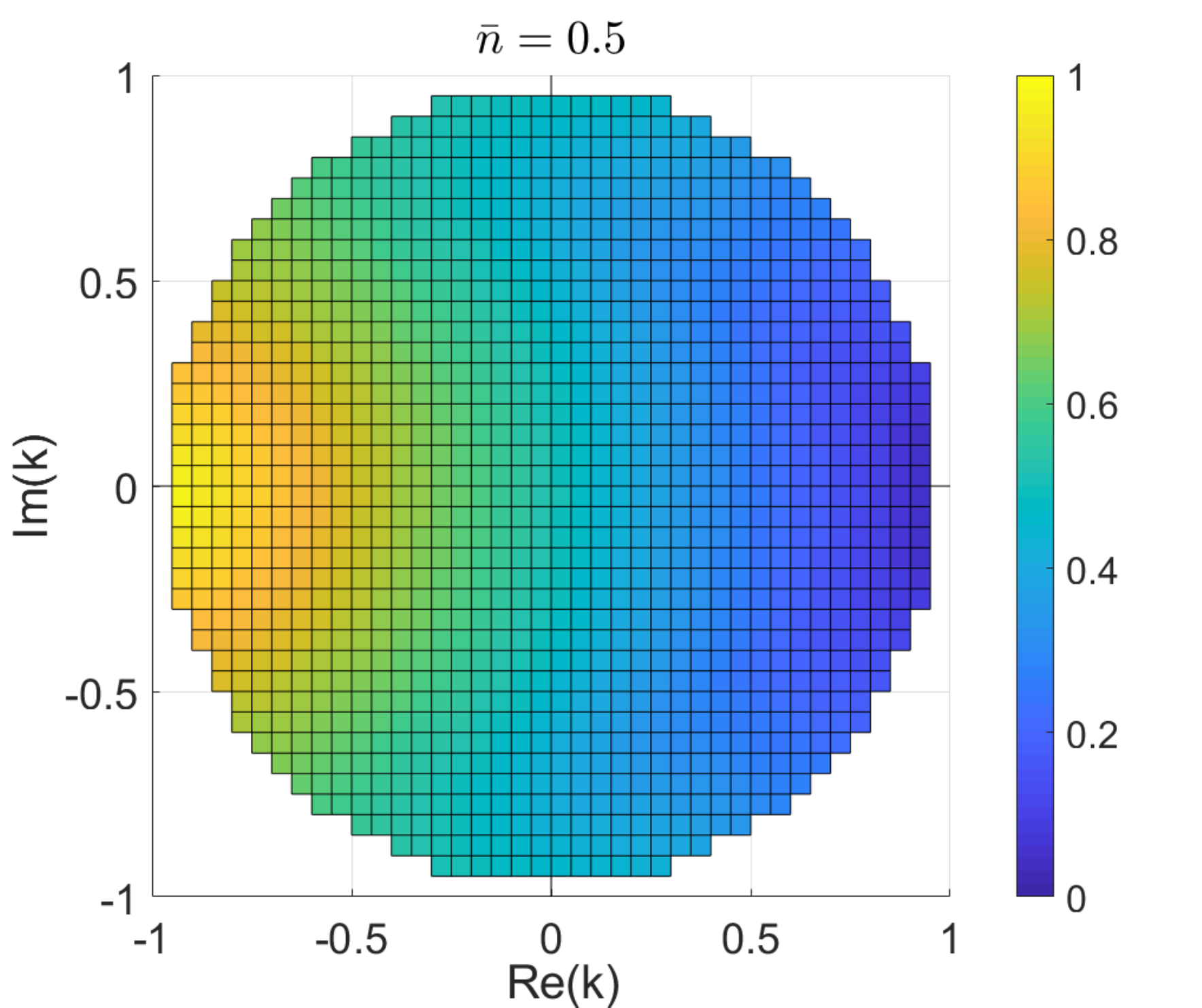}
        \caption{Upper bound $\pud^\text{SDP}$ for $\nbar = 0.5$}
        \label{fig: two_mode_UMD_nbar=0.5}
    \end{subfigure}
    \begin{subfigure}[b]{\columnwidth}
        \includegraphics[width= \columnwidth]{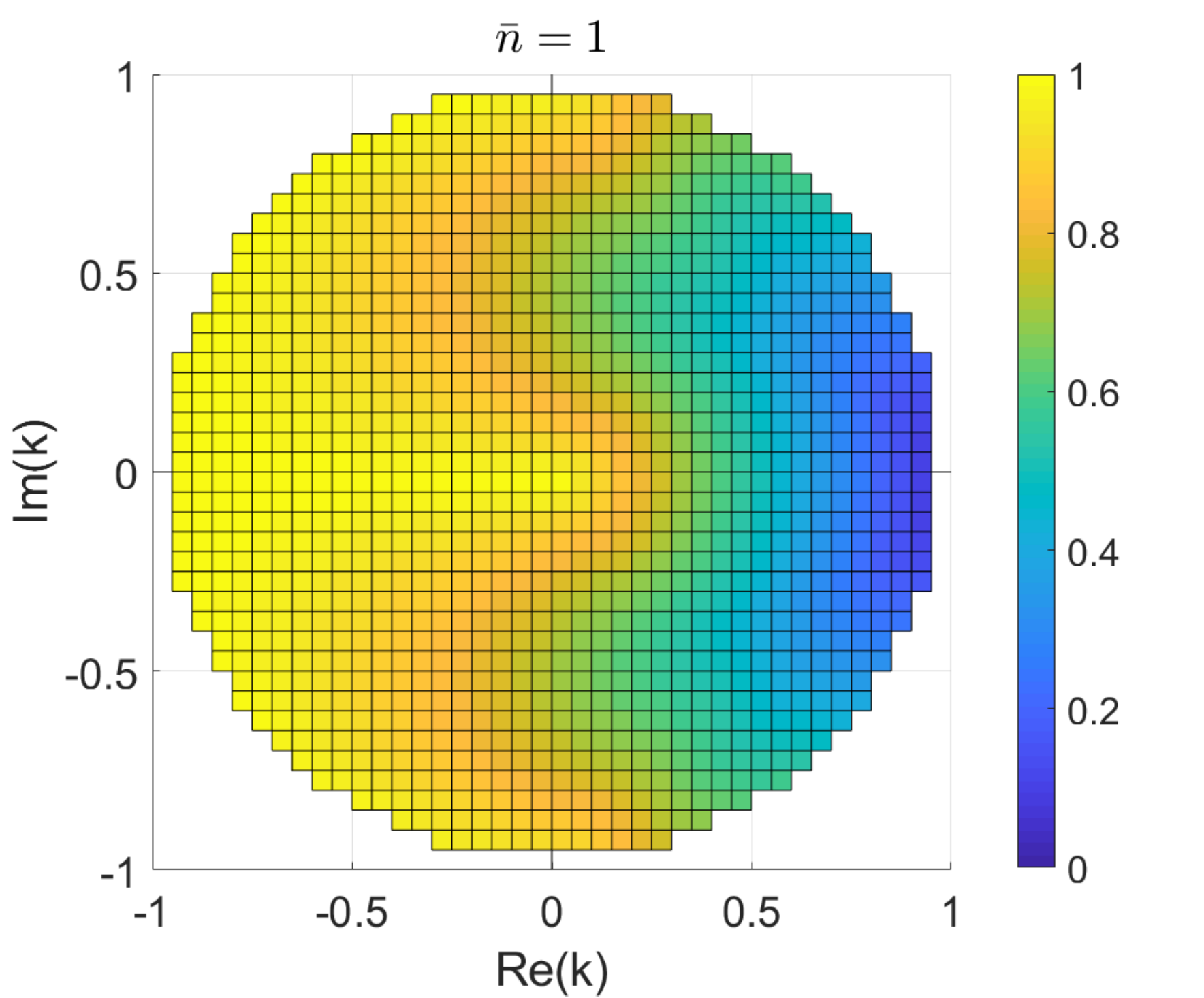}
        \caption{Upper bound $\pud^\text{SDP}$ for $\nbar = 1$}
        \label{fig: two_mode_UMD_nbar=1}
    \end{subfigure}
    \caption{\textbf{Unambiguous channel-discrimination between two modes: dependence on the mode overlap $k\in\mathbb{C}$}. Upper bound $\pud^\text{SDP}$ on the guessing probability, solution of the SDP \eqref{eq: SDP_UMD} for $\nmax = 300$, in a polar plot of $k$, for $\nbar = 0.5$ and $\nbar = 1$.} \label{fig: two_mode_UMD}
\end{figure}

We thus turn to our SDPs \eqref{eq: SDP_probabilistic} and \eqref{eq: SDP_UMD}. The results are shown in Fig.~\ref{fig: two_mode_prob} for probabilistic discrimination, and in Fig.~\ref{fig: two_mode_UMD} for unambiguous discrimination. As expected, the modes are harder to distinguish when $k \approx 1$. More remarkable is the fact that distinguishability improves significantly in the region of negative phases. For instance, while it is known and rather obvious that perfect discrimination for $k=0$ becomes possible for $\nbar\geq 1$, we see that when $k=-1$ the modes can already be perfectly distinguished for $\nbar=0.5$ (more in Section~\ref{sec: phase discrimination}).

\subsection{Phase discrimination}
\label{sec: phase discrimination}

Next we consider the problem of phase discrimination. Since the phase of an optical mode is not defined unless a reference beam is provided, this case study is restricted to the channel-discrimination scenario. The receiver's task is to guess one of a set of unitary channels $U_j = e^{i\varphi_j \hat{n}}$ where $\hat{n}$ is the number operator in the signal mode. In other words, the receiver can use pure states to distinguish modes of the form
\begin{equation}
    a_j = e^{i\varphi_j} a
\end{equation}
where $a$ is the initial signal mode prior to the phase-shift. 

\begin{figure}[H] 
    \centering
    \begin{subfigure}[b]{\columnwidth}
        \includegraphics[width= \columnwidth]{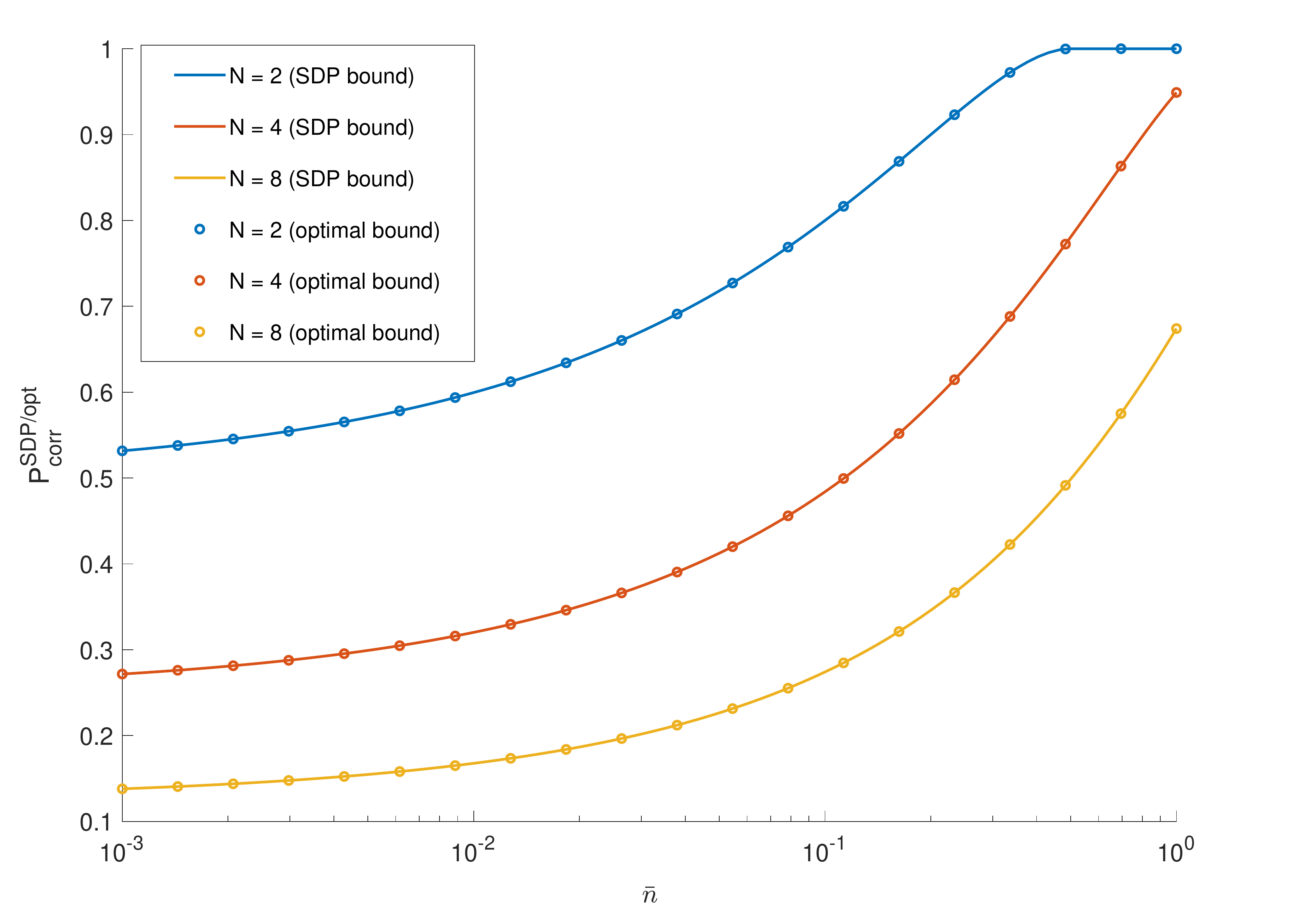}
        \caption{Optimal bound for probabilistic discrimination}
        \label{fig: MPSK_probabilistic}
    \end{subfigure}
    \begin{subfigure}[b]{\columnwidth}
        \includegraphics[width= \columnwidth]{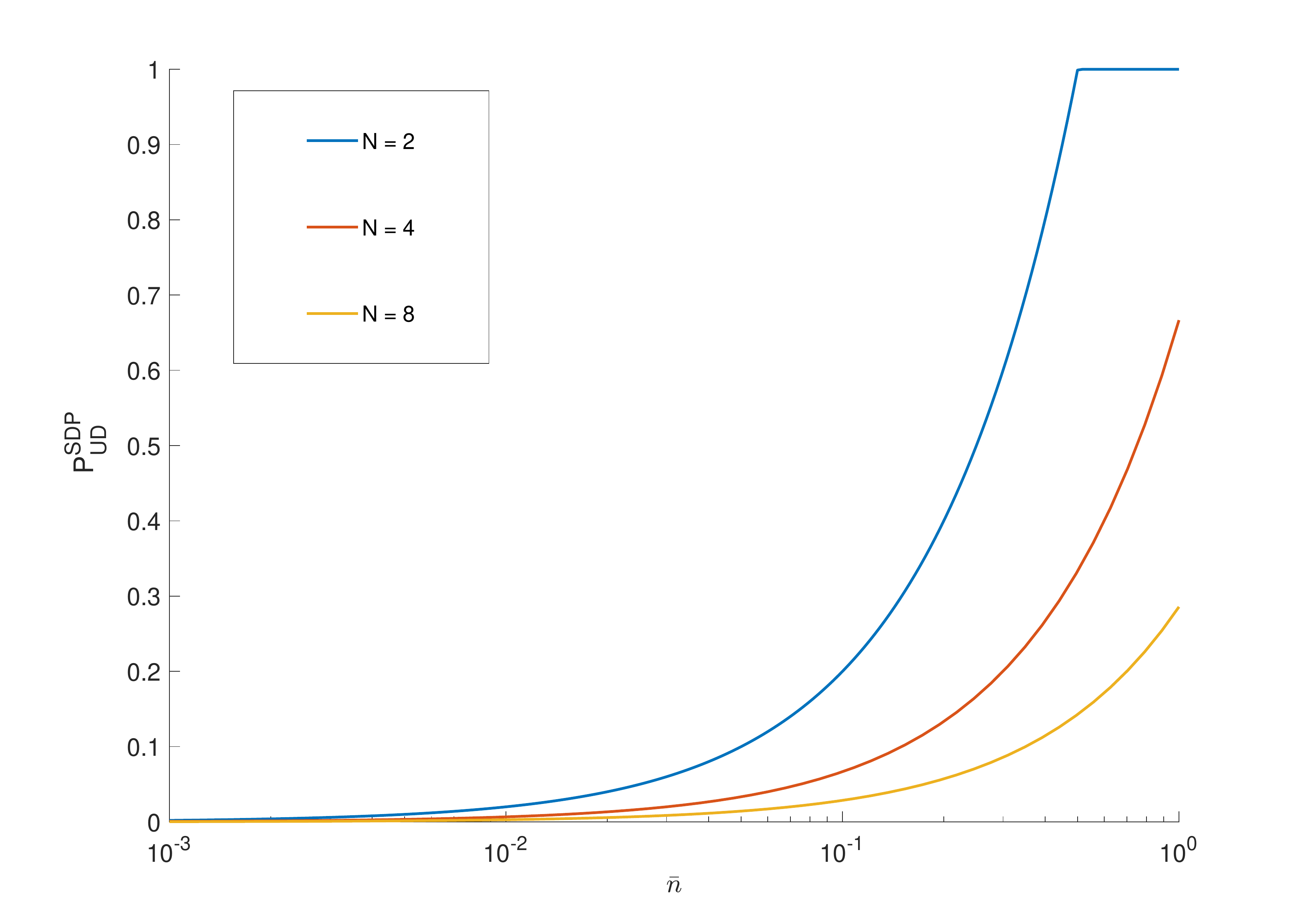}
        \caption{Optimal bound for unambiguous discrimination}
        \label{fig: MPSK_UMD}
    \end{subfigure}
    \caption{\textbf{Phase discrimination: dependence on $\nbar$}. (a) Upper bound $\pguess^\text{SDP}$ on the guessing probability, solution of the SDP \eqref{eq: SDP_probabilistic}. (b) Upper bound $\pud^\text{SDP}$ on the success probability, solution of the SDP \eqref{eq: SDP_UMD}. Both SDPs solved for $\nmax = 50$. In (a), the circles denote the analytical bounds from Theorem 3 of Ref.~\cite{nair2012}.} 
\end{figure}

Our formalism can be applied to \textit{any set of phases} to be discriminated. For the numerical case study, we choose the symmetric set $\{\varphi_j=2\pi j/\Nstate|j=0,...,\Nstate-1\}$, whose probabilistic discrimination has been studied in the context of quantum sensing ~\cite{nair2012}. The commutation relations are then given by
\begin{equation}
    [a_j, a_{k}^\dagger] = e^{i 2\pi (j-k) / \Nstate} \one\,.
\end{equation}
For probabilistic discrimination (see Fig.~\ref{fig: MPSK_probabilistic}), our SDP \eqref{eq: SDP_probabilistic} recovers the same bound as Theorem 3 of Ref.~\cite{nair2012}. The study of unambiguous discrimination is novel: the result is plotted in Fig.~\ref{fig: MPSK_UMD}. Also notice that, for $\Nstate > 2$, the modes in this family are linearly dependent; but, as discussed, unambiguous discrimination is possible because one can create linearly independent states.

The $N=2$ case corresponds to $k=-1$ in Section \ref{sec: two modes}. As we notice there and see again here, perfect discrimination becomes possible at $\nbar=0.5$. This is because the pure states with $c_0=c_1=\frac{1}{\sqrt{2}}$ are orthogonal. Indeed, $\ket{\psi_a}=\frac{1}{\sqrt{2}}(\ket{0}+a^\dagger\ket{0})=\frac{1}{\sqrt{2}}(\ket{0}+\ket{1}_a)$, while $\ket{\psi_{-a}}=\frac{1}{\sqrt{2}}(\ket{0}+(-a^\dagger)\ket{0})=\frac{1}{\sqrt{2}}(\ket{0}-\ket{1}_a)$. Since distinguishability cannot decrease when increasing $\nbar$, one expects to find two orthogonal states for any $\nbar\geq 0.5$. Indeed, these are
$\sqrt{\frac{1-\delta}{4}}\ket{m-1}_a\pm\frac{1}{\sqrt{2}}\ket{m}_a+\sqrt{\frac{1+\delta}{4}}\ket{m+1}_a$
with $\nbar=m+\delta/2$ and $-1\leq \delta<1$. In general these are superpositions of three Fock states, reducing to two only when $\nbar=m-\frac{1}{2}$ ($\delta=0$). Thus, as anticipated, the optimal state does not obey \eqref{floorceil}. 

\subsection{Computational and Fourier transform basis modes} \label{sec: FTbasis}
As our next example, we consider a family made of two sets of orthogonal modes. The first set $\mathcal{C}_d = \{ a_0, a_1, ..., a_{d-1} \}$ are called \textit{computational basis modes}. The second set $\mathcal{F}_d = \{b_0, b_1, ..., b_{d-1} \}$ are called \textit{Fourier transform basis modes} and are defined as
\begin{equation}
    b_k = \frac{1}{\sqrt{d}} \sum_{j = 0}^{d-1} (\omega_d)^{jk} a_j
\end{equation}
where $\omega_d = e^{2 \pi i/d}$ is the $d$-th root of unity. This family of modes have appeared in quantum cryptography: the classical information is encoded into these modes in the famous BB84 protocol \cite{bennett1984} for $d=2$, and the case $d>2$ defines one of its possible generalizations to higher alphabets \cite{sheridan2010security}. However, in those QKD protocols the classical information is encoded in the mode's \textit{index}: one wants to determine $j$, whether from $a_j$ or $b_j$. By contrast, here we stay with the task of mode discrimination, and study both probabilistic and unambiguous discrimination from the set $\mathcal{M}_d = \mathcal{C}_d \cup \mathcal{F}_d$.

\begin{figure}[H] 
    \centering
    \begin{subfigure}[b]{\columnwidth}
        \includegraphics[width= \columnwidth]{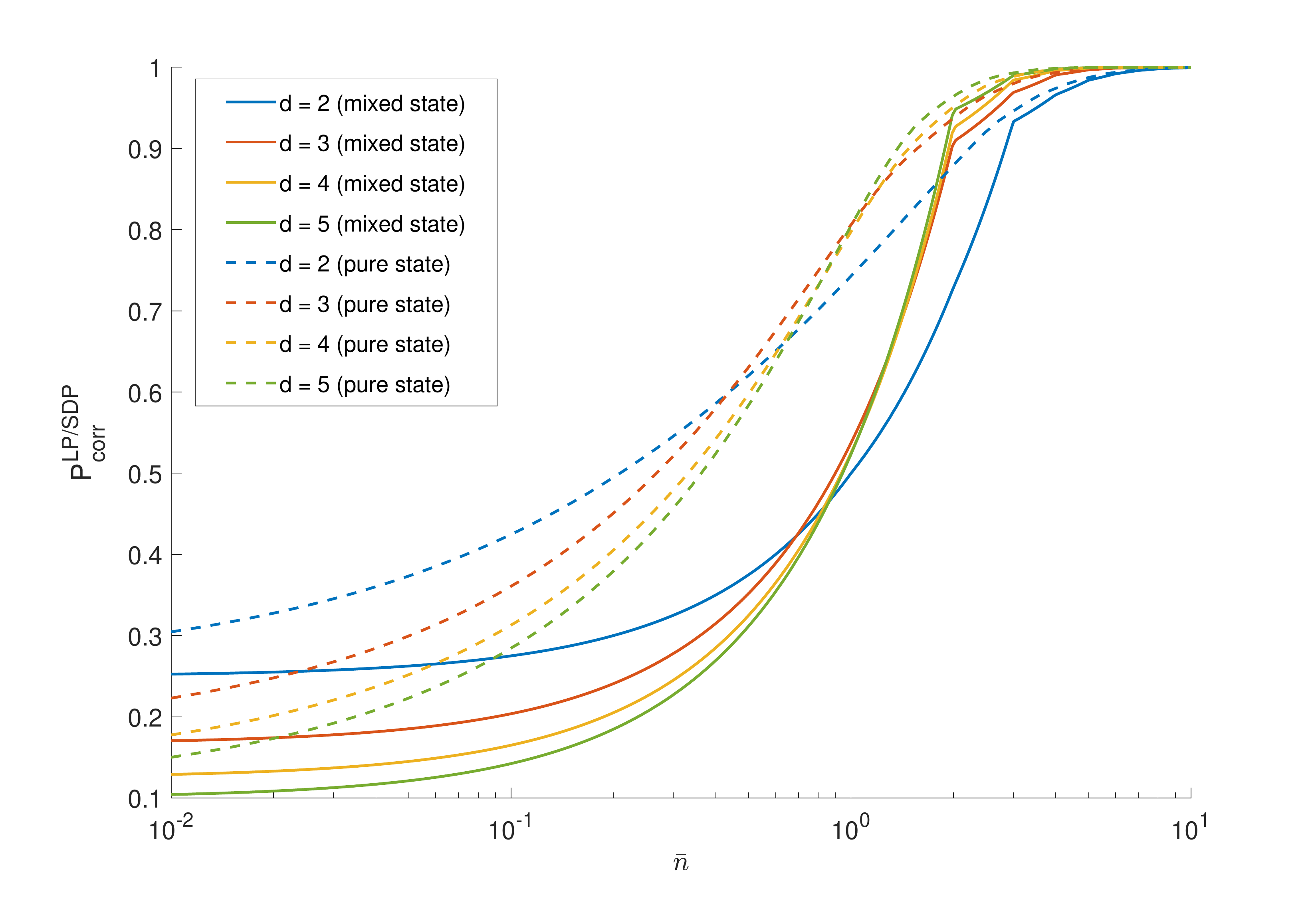}
        \caption{Optimal bound for probabilistic discrimination}
        \label{fig: computational_FT_probabilistic}
    \end{subfigure}
    \begin{subfigure}[b]{\columnwidth}
        \includegraphics[width= \columnwidth]{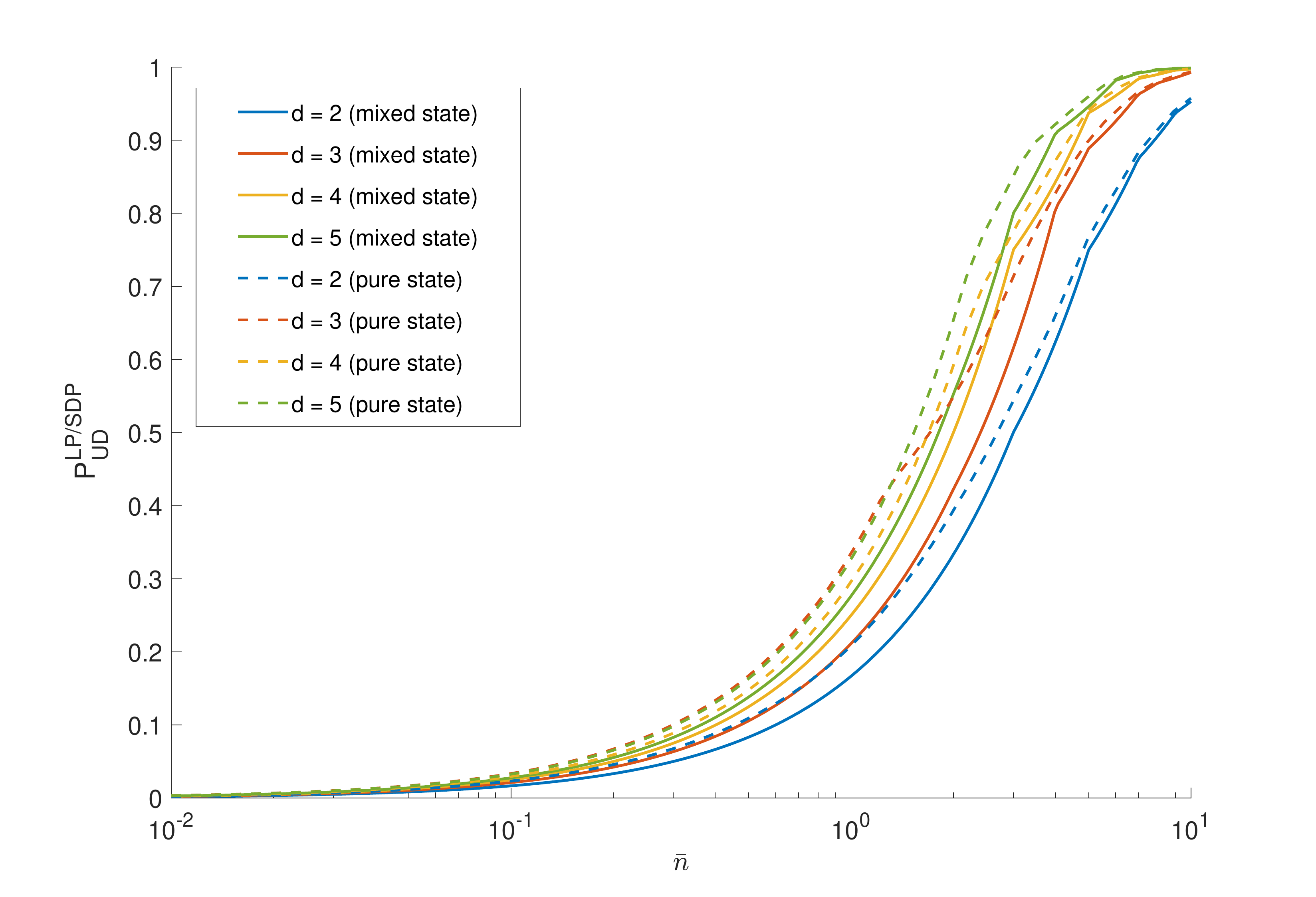}
        \caption{Optimal bound for unambiguous discrimination}
        \label{fig: computational_FT_UMD}
    \end{subfigure}
    \caption{\textbf{Computational and Fourier transform basis modes: dependence on $\nbar$}. (a) Upper bound on the guessing probability for probabilistic discrimination. (b) Upper bound on the success probability for unambiguous discrimination. The solid lines are the bounds in the source-discrimination scenario, obtained by solving the LP \eqref{eq: LP pguess relaxed} or \eqref{eq: LP pud relaxed} respectively whereas the dashed curves the bounds in the channel-discrimination scenario obtained by solving the SDP \eqref{eq: SDP_probabilistic} or \eqref{eq: SDP_UMD}. For this family of modes, it is sufficient to set $\nmax = 50$.} 
\end{figure}

For given $d$, the commutation relations are
\begin{equation}
    \begin{split}
        [a_j, a_k^\dagger] &= \delta_{jk} \one\\
        [b_j, b_k^\dagger] &= \delta_{jk} \one\\
        [a_j, b_k^\dagger] &= \frac{1}{\sqrt{d}} \left(\omega_d^{*} \right)^{jk} \one\\
    \end{split}
\end{equation} 
We set $\nmax = 50$ and solve the SDP and/or LP for $\bar{n}$ varying from $10^{-3}$ to 10, and for $d=2,3,4,5$. The results for probabilistic discrimination are shown in Fig. \ref{fig: computational_FT_probabilistic}; those for unambiguous discrimination in \ref{fig: computational_FT_UMD}; both figures contain information about the channel-discrimination scenario (dashed) and the source-discrimination scenario (solid).

Channel discrimination is more powerful than source discrimination: while more marked for probabilistic discrimination, the difference is also present in unambiguous discrimination, contrary to what was the case for two modes (Section \ref{sec: two modes}). Another feature present in both figures, again more marked for probabilistic discrimination, is a crossover of behavior as a function of $\nbar$: for $\nbar\lesssim 1$, the discrimination is better for smaller $d$; whereas for $\nbar\gtrsim 1$, the discrimination is better for higher $d$. This can be understood qualitatively. In the limit $\nbar\rightarrow 0$, guessing the mode is hardly more than a random pick from a uniform distribution, so the guessing probability is close to $\frac{1}{2d}$ and decreases with $d$. In the limit of large $n$, $|\braket{n_{a_j}}{n_{b_k}}|=\sqrt{1/d^n}$ decreases with $d$ and therefore the distinguishability increases.

In the source-discrimination scenario, we find that condition \eqref{eq: condition} is generally not satisfied for $\nbar=1$, either for probabilistic or unambiguous discrimination. 

For unambiguous discrimination, the reason is clear: the single-photon states are linearly dependent and hence unambiguous discrimination is not possible. In order for unambiguous discrimination to be possible, the state must contain some multi-photon component.

\begin{figure}[H]
    \centering
    \includegraphics[width=\columnwidth]{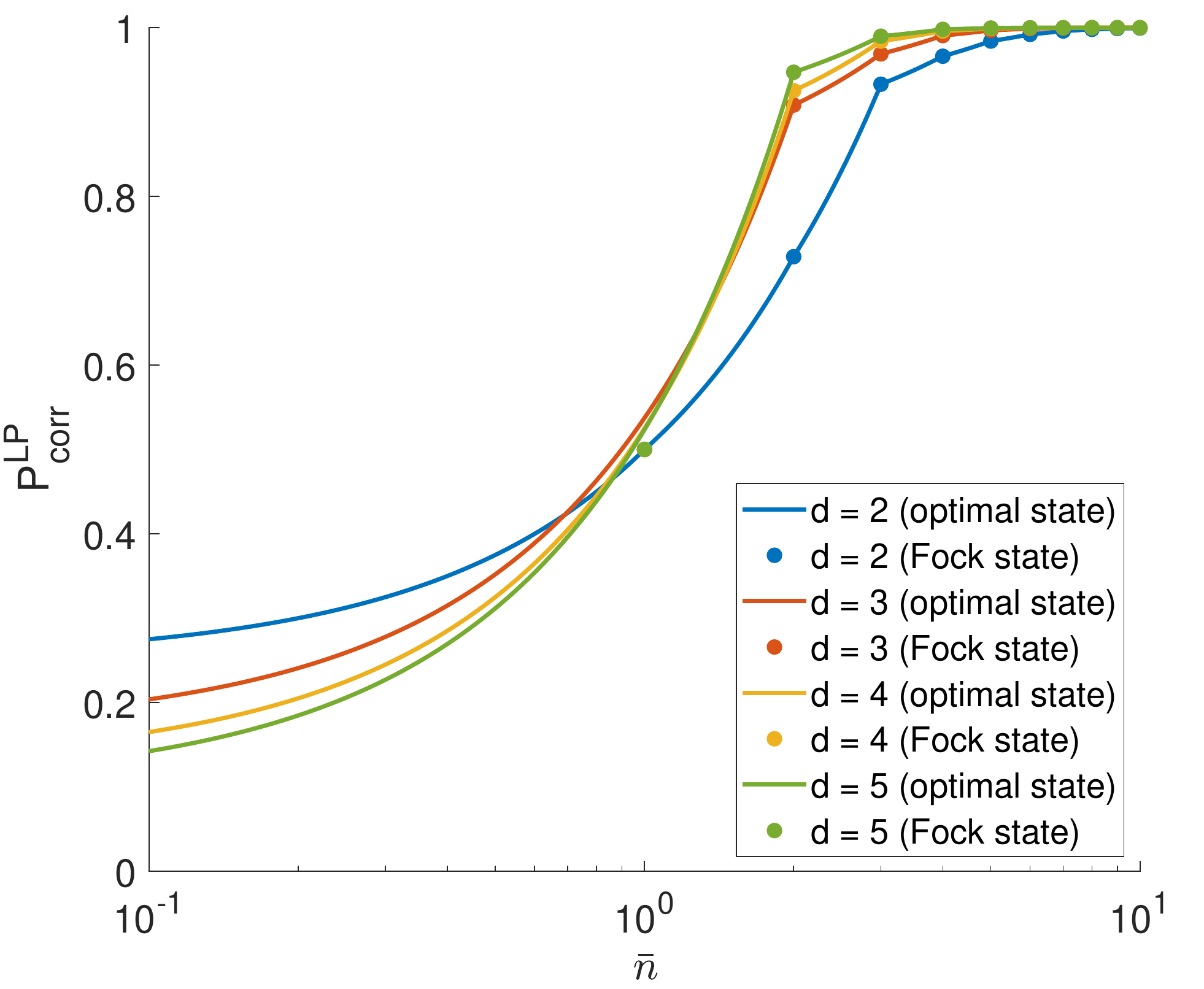}
    \caption{\textbf{Computational and Fourier transform basis modes: optimal state vs Fock states (for probabilistic discrimination in the source-discrimination scenario)}. The solid lines are the bound $\pguess^{\text{LP}}$, the solution to \eqref{eq: LP pguess relaxed} with $\nmax = 50$. On the other hand, the dots represent the bound $\pguess^{(n)}$, the solution to the SDP \eqref{eq: pguess Fock state}.}
    \label{fig: optimal vs Fock computationalFT}
\end{figure}

For probabilistic discrimination, we find that the condition \eqref{eq: condition} is violated for $d=3,4,5$. To see that, we compare $\pguess^{\text{LP}}$ with that of Fock states in Fig.~\ref{fig: optimal vs Fock computationalFT}. For $d = 2,3,4,5$, the single-photon bound is $\pguess^{(1)} = 0.5$. This corresponds to a simple strategy: betting on one of the bases and measure in it.

\subsection{Differential-phase-shift modes} \label{sec: DPS}
Lastly, we consider a family of modes inspired by the differential-phase-shift (DPS) QKD protocol \cite{inoue2002differential}. In the protocol, the information is encoded in the relative phase between subsequent temporal modes. Abstractly, to any $\ell$-bit strings $\mathbf{x}$, a mode is associated according to
\begin{equation}
    b_{\mathbf{x}} = \frac{1}{\sqrt{\ell + 1}}\left( a_0 + \sum_{i = 1}^{\ell} e^{i \varphi_i^{(\mathbf{x})}} a_i \right),
\end{equation}
where the $\{ a_i \}_{i=0,...,\ell}$ are $(\ell + 1)$ orthogonal modes (temporal ones in the original setting) and where
\begin{equation} \label{eq: differential phase}
    \varphi_i^{(\mathbf{x})} - \varphi_{i-1}^{(\mathbf{x})} = x_i \pi
\end{equation}
with $x_i \in \{0,1\}$ the $i$-th bit of the string $\mathbf{x}$ (by convention, we set the phase of the reference mode $\varphi_0^{(\mathbf{x})} = 0$ for all $\mathbf{x}$). For a given $\ell$, there are $2^\ell$ different modes, only linearly-many of which are orthogonal among the $2^\ell$ ones. The commutation relation for the modes associated to string $\mathbf{x}$ and $\mathbf{y}$ can be computed recursively using the relation \eqref{eq: differential phase} and is given by
\begin{equation}
    \left[b_{\mathbf{x}}, b^\dagger_{\mathbf{y}} \right] = \frac{1}{\ell+1} \left( 1 + \sum_{i = 1}^{\ell} e^{i(\varphi_i^{(\mathbf{x})} - \varphi_i^{(\mathbf{y})})} \right)\one\,.
\end{equation}

Since the SDP scales badly with $\ell$, in this paper we present the results only for $\ell=1,2,3$. When $\ell = 1$, the two modes to be distinguished are orthogonal, and so this is a special case of what we studied in Section \ref{sec: two modes}. For $\ell=2$, the four modes are all non-orthogonal. The eight modes for $\ell=3$ form two sets of four orthogonal modes. 

\begin{figure}[H] 
    \centering
    \begin{subfigure}[b]{\columnwidth}
        \includegraphics[width= \columnwidth]{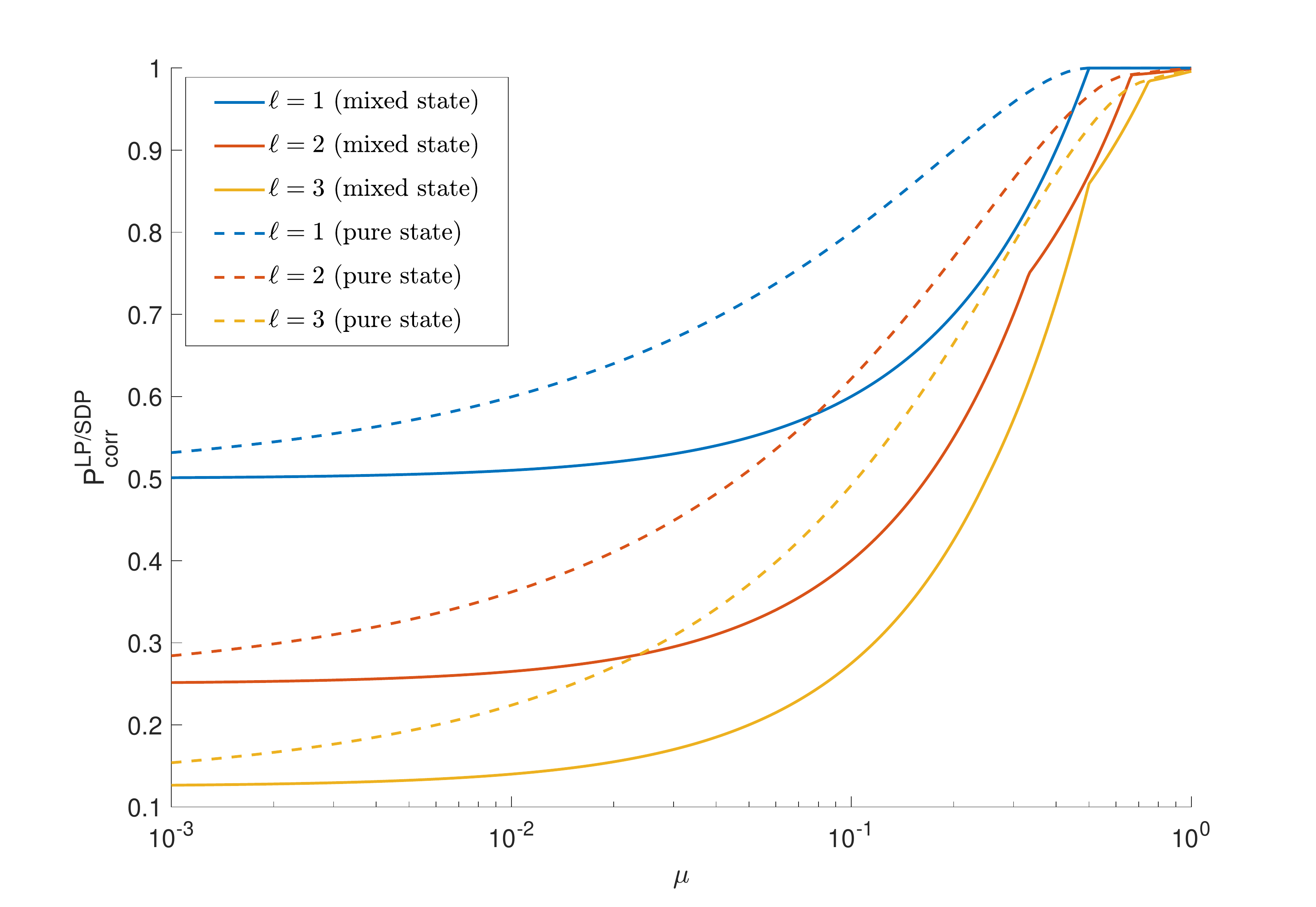}
        \caption{Optimal bound for probabilistic discrimination}
        \label{fig: DPS_probabilistic}
    \end{subfigure}
    \begin{subfigure}[b]{\columnwidth}
        \includegraphics[width= \columnwidth]{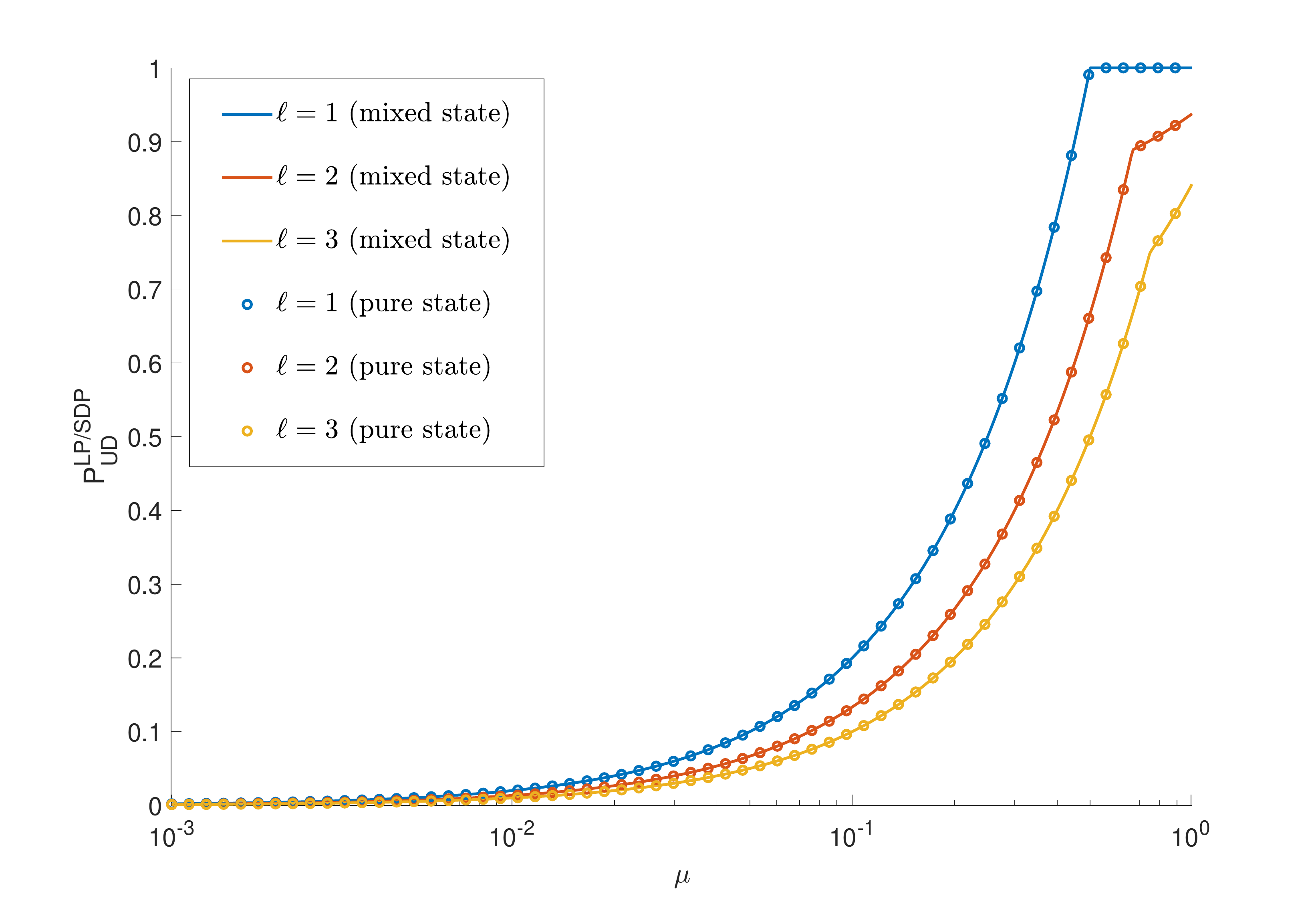}
        \caption{Optimal bound for unambiguous discrimination}
        \label{fig: DPS_UMD}
    \end{subfigure}
    \caption{\textbf{Discriminating the differential-phase-shift modes: dependence on the energy per pulse $\mu$}. (a)~Upper bound on the guessing probability for probabilistic discrimination. (b)~Upper bound on the success probability for unambiguous discrimination. In both cases, the solid lines are the bounds for source-discrimination scenario: solutions to the LPs \eqref{eq: LP pguess relaxed} and \eqref{eq: LP pud relaxed} respectively. The dashed curves in (a) and the circles in (b) are the bounds for channel-discrimination scenario obtained by solving the SDPs \eqref{eq: SDP_probabilistic} and \eqref{eq: SDP_UMD} respectively. Here, we set $\nmax = 50$.} 
\end{figure}

The results of our numerical method are shown in Fig.~\ref{fig: DPS_probabilistic} for probabilistic discrimination, and in Fig.~\ref{fig: DPS_UMD} for unambiguous discrimination. Since the family given by $\ell$ is constructed from $\ell+1$ orthogonal modes (``pulses''), we found it more appropriate to compare the families for a given value of the \textit{energy per pulse} $\mu = \nbar/(\ell+1)$, rather than of the total energy $\nbar$. Even with this scaling, it proves more difficult to distinguish within a set with higher $\ell$, since the receiver has to discriminate more modes.

Comparing the mixed state encoding to the pure state encoding, we find that the pure state encoding is more distinguishable only in the probabilistic discrimination setting. Remarkably, we find that the mixed state bounds coincide with that of the pure state encoding for the unambiguous discrimination setting.

\section{Coda: on losses}
\label{sec:losses}

\begin{figure}[H]
    \centering
    \includegraphics[width= 1 \columnwidth]{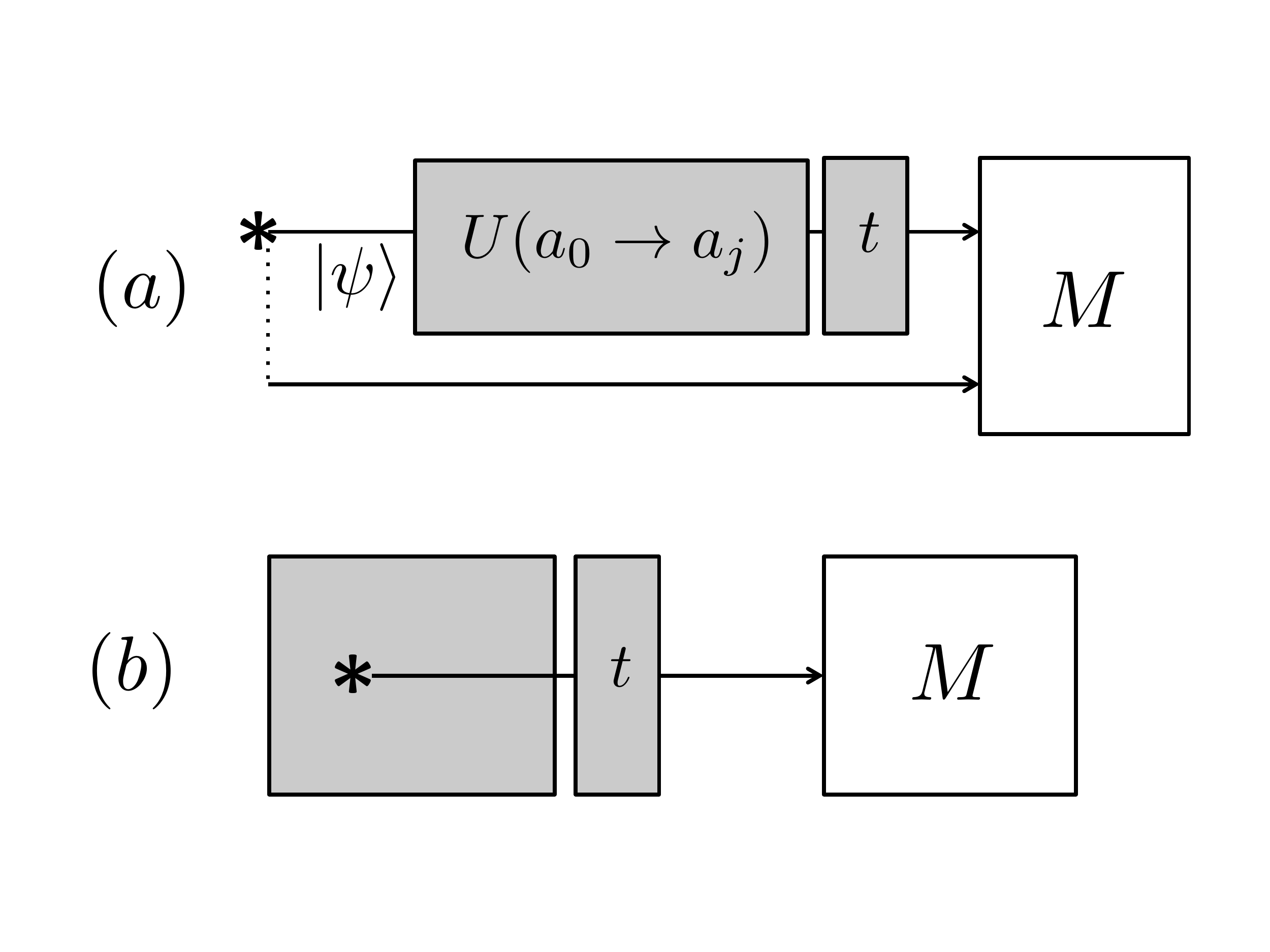}
    \caption{The two scenarios studied in this paper, with additional mode-independent losses before the measurement device ($t$ refers to the transmittivity).}
    \label{figlossy}
\end{figure}

In this last Section, we review how the ultimate limits of mode discrimination are modified when there are \textit{mode-independent losses} between the device to be tested and the measurement, as sketched in Fig.~\ref{figlossy}. This is modelled as a beam splitter transformation $a_{j} \longrightarrow t a_{j}+r b_{j}$ where $t$ and $r$ can be taken as real and positive and $t^2 + r^2 = 1$. The state that arrives at the measurement station is now the partial state associated with the transmitted output.

The study of the source-discrimination scenario remains practically unchanged. Indeed, losses do not introduce any coherence between photon number states. So, if the state before the losses is given by Eq. \eqref{eq: mixed state}, the state after the losses will still be a number mixture $\rho_j(t)=\sum_{n = 0}^\infty q_n(t) \ketbra{n_j}{n_j}$ with
\ba\label{qn}
q_n(t)&=& \sum_{m \geq n} p_m \binom{m}{n} (1-t^2)^{m-n} t^{2n}\,. 
\ea If $\nbar$ still represents the energy constraint for the input state, one solves the LP under the energy constraints $\nbar t^2$ to get the $\{q_n(t)\}$. The $\{p_n\}$ to be prepared are then obtained by inverting the linear system of equations \eqref{qn}. The only practical worries may come from numerical cutoffs in this inversion when $\nbar$ is large.

On the contrary, the optimizations in the channel-discrimination scenario can no longer be cast as SDPs. Indeed, the initially pure state becomes mixed with losses. Thus, on the one hand, we cannot build Gram matrices as we did in Section \ref{sec:channel}. But on the other hand, the basis in which the mixed state is diagonal is not fixed, and is definitely not the Fock basis: so we cannot adapt the approach of Section \ref{sec:source} either.

In some simple cases, a \textit{heuristic} optimization may still be trustful. For instance, let us look at the probabilistic discrimination of two modes with $k\geq 0$. We know \cite{helstrom1969quantum} that the probability of discriminating correctly between two equally probable mixed states is given by $\pguess(\rho_1,\rho_2)=\frac{1}{2}\left(1+\frac{1}{2}\textrm{Tr} |\rho_1-\rho_2|\right)$. One can then write the algorithm that, given a $\ket{\psi_j}$ in the form \eqref{eq: pure states}, computes the $\rho_j(t)$ obtained after losses; then heuristically maximizes the trace distance $\left|\rho_1(t)-\rho_2(t)\right|$ over the complex parameters $c_n$ under the energy constraint $\sum_n n|c_n|^2=\nbar$ for the initial states. 
We implemented this procedure using the function fmincon of MATLAB. Inspection of the numerical solutions we obtained indicates that the relevant parameter in the input state are the $p_n=|c_n|^2$, while the arguments of the $c_n$ (relative phases) do not seem to matter, just as in the lossless case \eqref{eq: minscal}.

\begin{figure}[H]
    \centering
    \includegraphics[width=\columnwidth]{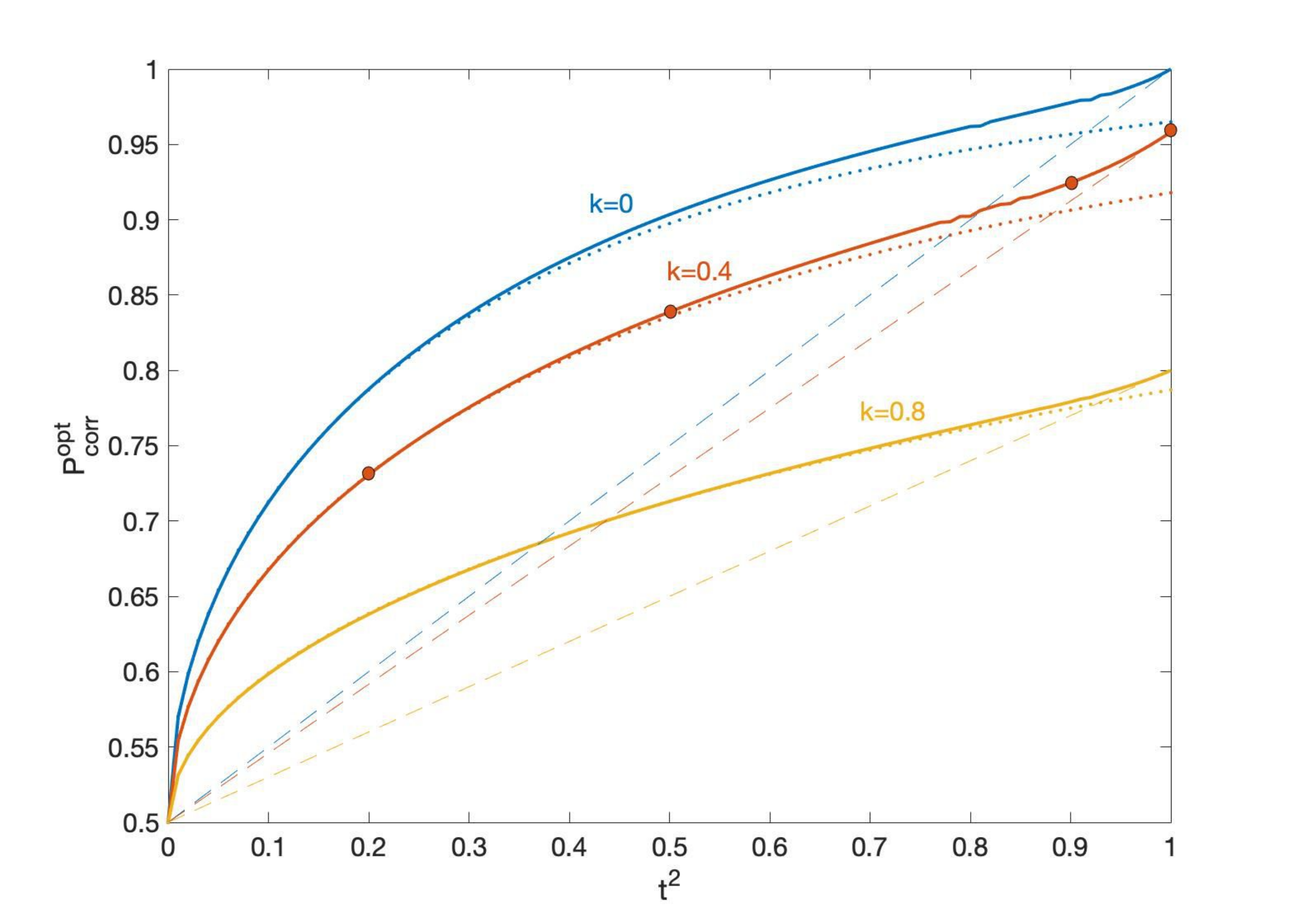}
    \caption{\textbf{Probabilistic discrimination of two modes as a function of losses (channel-discrimination scenario).} Plot of $P_{\text{corr}}^{\text{opt}}(k,\bar{n},t)$ as a function of $t^2$ for $\bar{n}=1$ and $k=0, 0.4, 0.8$ (from top to bottom). The thick solid lines are the result of the heuristic optimisation (for which the photon number was truncated at 5); on the line for $k=0.4$ are indicated the points reported in Table \ref{weights}. The dotted lines are the values $P_{\text{corr}}^\textrm{{coh}}$ for coherent state input; the dashed lines are the values $P_{\text{corr}}^\textrm{{Fock}}$ for Fock state input.}
    \label{asaphlosses}
\end{figure}

An example is given in Fig.~\ref{asaphlosses}. The constraint is set at $\bar{n}=1$: thus, in the absence of losses, the optimal state is the Fock state $\ket{1}$. When losses increase, this state becomes quickly sub-optimal, while the probability $P_{\text{corr}}^{\text{opt}}(k,\bar{n},t)$ of guessing correctly approaches the value corresponding to choosing a \textit{coherent state} as input state (see also Table \ref{weights}). Based on this observation, whenever $\bar{n}=m$ integer, for the sake of estimates one could use
\ba\label{estimate}
P_{\text{corr}}^{\text{opt}}&\gtrsim&\max\left(P_{\text{corr}}^{\text{coh}},P_{\text{corr}}^{\text{Fock}}\right)
\ea because the two probabilities on the r.h.s.~can be given analytically. Indeed, on a beam-splitter, a coherent state splits as $\ket{\alpha}\rightarrow\ket{t\alpha}_T\ket{r\alpha}_R$; so the states in the transmitted mode are pure, and a standard calculation gives $\bra{t\alpha}_0\ket{t\alpha}_1=e^{-|t\alpha|^2(1-k)}$, and finally $P_{\text{corr}}^{\text{coh}}(k,\bar{n},t) = \frac{1}{2}(1+\sqrt{1- e^{-2t^2\,\bar{n}(1-k)}})$.
For a $m$-photon Fock state, the transmitted state reads $\rho_T=\sum_{n=0}^m q(n,m,t)\ket{n}\bra{n}$ with $q(n,m,t)={m\choose n}t^{2n}r^{2(m-n)}$. Because it's diagonal in the number basis, the optimal measurement to discriminate between the modes can be seen as: first, measure the photon number and thus project in a Fock state; then distinguish between the two Fock states. Thus $P_{\text{corr}}^\textrm{{Fock}}(k,m,t) = \frac{1}{2}(1+\sum_{n=0}^m q(n,m,t)\sqrt{1- k^{2n}})$.

\begin{table}[H]
\centering
    \begin{tabular}{|c||c|c|c|c|c|c|}
    \hline
         $t^2$ & $p_0$ & $p_1$ & $p_2$ & $p_3$ & $p_4$ & $p_5$ \\
         \hline
        1 (Fock) & 0 & 1 & 0 & 0 & 0 & 0\\
        \hline
        0.9 & 0.1458 & 0.7088 & 0.1454 & 0.0000 & 0.0000 & 0.0000 \\
        \hline
        0.5 & 0.3075 & 0.4363 & 0.2092 & 0.0430 & 0.0039 & 0.0001  \\
        \hline
        0.2 & 0.3450 & 0.3914 & 0.1952 & 0.0567 & 0.0105 & 0.0012  \\
        \hline
        coh & \textit{0.3679} & \textit{0.3679} & \textit{0.1839} & \textit{0.0613} & \textit{0.0153} & \textit{0.0031}  \\
        \hline
    \end{tabular}
    \caption{Photon-number weights of the optimal input states, for $\bar{n}=1$ and $k=0.4$ (thick dots in Fig.~\ref{asaphlosses}). The result for $t^2=1$ is analytical [see \eqref{eq: chi} in Section \ref{sec: two modes}], the other values are the output of the heuristic optimization. The coherent state weights $e^{-1}/n!$ are given for reference in the last line.} \label{weights}
\end{table}

\section{Conclusion}

In this paper, we have presented efficient methods to compute the ultimate limits for single-shot discrimination of optical modes. The methods, based on linear and semi-definite programming, apply to any set of modes with any prior distribution (we wrote the paper for the uniform prior not to introduce further notation, but the modifications are obvious). The bounds that are obtained can be used as fundamental benchmark for the performance of realistic devices or measurement schemes.

We pointed out the importance of stating whether the verifier has the possibility of defining a reference frame for the modes' phase. Depending on the family of modes that is considered, the difference in discrimination is found to be significant; and of course, some tasks like phase discrimination make only sense if the reference is available. Note that we assumed that one the reference beam is classical and hence its phase relative to the receiver's reference frame could be determined with arbitrary precision. We leave the study of the channel-discrimination scenario with a weak reference beam as an open problem.

Let us finish by pointing out two related topics that we have not dealt with in the current work. First: throughout the paper, the characterization of the optical modes has been taken as known and perfect. It is known that this could be relaxed in some situations. Indeed, randomness of quantum origin can be certified from the measurement of uncharacterized optical modes, based only on an energy constraint $\nbar<1$ \cite{vH16}. Second: we have considered single-shot discrimination. For the discrimination of unitaries, it is known that perfect discrimination is always possible if one has enough many copies \cite{acin01}, and a similar result for energy-constrained discrimination has been described recently \cite{datta2020}.

\section*{Acknowledgments}

We thank Chee Wei Soh for help in the first part of this project, and Konrad Banaszek, Alessandro Bisio, Michele Dall'Arno, Claude Fabre, Charles Lim, Norbert L\"utkenhaus, Matteo Paris and Stefano Pirandola for comments and suggestions. This research is supported by the National Research Foundation and the Ministry of Education, Singapore, under the Research Centres of Excellence programme.

\bibliography{References}
\end{document}